\documentclass{elsarticle}

\usepackage{lineno,hyperref}
\modulolinenumbers[5]

\usepackage{amssymb}
\usepackage{amsmath}
\usepackage{stmaryrd}
\usepackage{multirow}
\usepackage{subfig}
\usepackage{bm}
\usepackage{ulem,cancel,soul}
\usepackage{listings}
\usepackage{color}
\usepackage{hyperref}

\usepackage{pdfpages}
\usepackage{graphicx}
\usepackage{blindtext}

\usepackage{xcolor}
\usepackage{tikz}
\colorlet{MyColorOne}{blue!50}
\usetikzlibrary{shapes.geometric, arrows}
\tikzstyle{startstop} = [rectangle, rounded corners, minimum width=6cm, minimum height=1cm,text centered, draw=black]

\tikzstyle{startstop1} = [rectangle, rounded corners, minimum width=10cm, minimum height=1cm,text centered, draw=black]

\tikzstyle{arrow} = [thick,->,>=stealth]

\newcommand{\lightercolor}[3]{
    \colorlet{#3}{#1!#2!white}
}

\newcommand{\darkercolor}[3]{
    \colorlet{#3}{#1!#2!black}
}

\lightercolor{MyColorOne}{50}{MyColorOneLight}
\darkercolor{MyColorOne}{50}{MyColorOneDark}

\usepackage{sistyle}
\SIthousandsep{,}

\definecolor{dkgreen}{rgb}{0,0.6,0}
\definecolor{gray}{rgb}{0.5,0.5,0.5}
\definecolor{mauve}{rgb}{0.58,0,0.82}

\newcommand{\ben}{\begin{equation*}}
\newcommand{\een}{\end{equation*}}
\newcommand{\be} [1] {\begin{equation} \label{#1}}
\newcommand{\ee}{\end{equation}}

\newcommand{\bx}{\boldsymbol{x}}

\newcommand{\BE}{\begin{equation}}
\newcommand{\EE}{\end{equation}}

\newcommand{\bunit}{\ensuremath{\text{s/mm}^2}}
\newcommand{\dunit}{\ensuremath{\text{mm}^2/\text{s}}}
\newcommand{\kunit}{\ensuremath{\text{m}/\text{s}}}
\newcommand{\lunit}{\ensuremath{\mu\text{m}}}
\newcommand{\tunit}{\ensuremath{\mu\text{s}}}

\newcommand\irregularcircle[2]{
  \pgfextra {\pgfmathsetmacro\len{(#1)+rand*(#2)}}
  +(0:\len pt)
  \foreach \a in {10,20,...,350}{
    \pgfextra {\pgfmathsetmacro\len{(#1)+rand*(#2)}}
    -- +(\a:\len pt)
  } -- cycle
}

\journal{Elsevier}

\begin{document}

\begin{frontmatter}

\title{Portable simulation framework for diffusion MRI}

\author[label1]{Van-Dang Nguyen \corref{cor1}}
\address[label1]{Division of Computational Science and Technology, KTH Royal Institute of Technology, Sweden}
\ead{vdnguyen@kth.se}

\author[label1]{Massimiliano Leoni} 
\author[label4,label1]{Tamara Dancheva} 
\address[label4]{Basque Center for Applied Mathematics (BCAM), Bilbao, Spain.}

\author[label1]{Johan Jansson}
\ead{jjan@kth.se}

\author[label1]{Johan Hoffman}
\ead{jhoffman@kth.se}

\author[label3]{Demian Wassermann}
\ead{demian.wassermann@inria.fr}
\address[label3]{Parietal, INRIA, Paris, France}

\author[label2]{Jing-Rebecca Li}
\ead{jingrebecca.li@inria.fr}
\address[label2]{INRIA Saclay-Equipe DEFI, CMAP, Ecole Polytechnique
Route de Saclay, 91128, Palaiseau Cedex, France}

\cortext[cor1]{I am corresponding author}

%
%
%

\begin{abstract}
The numerical simulation of the diffusion MRI signal arising from complex tissue micro-structures
is helpful for understanding and interpreting imaging data as well as for designing and optimizing MRI sequences.
The discretization of the Bloch-Torrey equation by finite elements is a more recently developed approach 
for this purpose, in contrast to random walk simulations, which has a longer history.  
While finite elements discretization is more difficult to implement than random walk simulations, 
the approach benefits from a long history of theoretical and numerical developments 
by the mathematical and engineering communities.  
In particular, software packages for the automated solutions of partial differential equations using finite elements discretization, such as FEniCS, are undergoing active support and development. 
However, because diffusion MRI simulation is a relatively new application area, there is still a gap 
between the simulation needs of the MRI community and the available tools provided by 
finite elements software packages.  In this paper, we address two potential difficulties 
in using FEniCS for diffusion MRI simulation. First, we simplified software installation by  the use of FEniCS containers that are completely portable across multiple platforms. 
Second, we provide a portable simulation framework based on Python and whose code is open source.  
This simulation framework can be seamlessly integrated with cloud computing resources such as 
Google Colaboratory notebooks working on a web browser or with Google Cloud Platform with MPI parallelization. 
We show examples illustrating the accuracy, the computational
times, and parallel computing capabilities. The framework contributes to reproducible science and open-source software in computational diffusion MRI with the hope that it will help to speed up method developments and stimulate research collaborations.

\end{abstract}


\end{frontmatter}


\section{Introduction}

The numerical simulation of the diffusion MRI signal arising from complex tissue micro-structures
is helpful for understanding and interpreting imaging data as well as for designing and optimizing MRI sequences. 

It can be classified into two main groups. The first group is referred to as Monte-Carlo simulations in the literature and previous works include \cite{nla.cat-vn2111911, Yeh2013, 4797853,Palombo2016, VANNGUYEN2018}.   
Software packages include the UCL Camino Diffusion MRI Toolkit \cite{Cook2006CaminoOD},
which has been widely used in the field. 
The second group of simulations relies on solving the Bloch-Torrey PDE in a geometrical domain, 
either using finite difference methods (FDM) \cite{Hwang2003,Xu2007,Harkins2009,Russell2012}, typically on a Cartesian grid, or finite element methods (FEM), typically on a tetrahedral grid. 
Previous works on FEM include \cite{Moroney2013} for the short gradient pulse limit of some simple geometries,
\cite{Nguyen2014283} for the multi-compartment Bloch-Torrey equation with general gradient pulses, and \cite{BELTRACHINI2015126} with the flow and relaxation terms added. 
In \cite{1742-6596-490-1-012013}, a simplified 1D manifold Bloch-Torrey equation was solved to 
study the diffusion MRI signal from neuronal dendrite trees. 
FEM in a high-performance computing framework was proposed in \cite{Nguyen1080573, NGUYEN2018271} for 
diffusion MRI simulations on supercomputers.  An efficient simulation method for thin media was proposed in \cite{NGUYEN2019176}.
A comparison of the Monte-Carlo approach with the FEM approach for the short pulse limit was performed in \cite{Moroney2013}, where FEM simulations were evaluated to be more accurate and faster than the equivalent modeling with Monte-Carlo simulations.
Recently, SpinDoctor, a Matlab-based diffusion MRI simulation toolbox that discretizes the Bloch-Torrey equation using 
finite elements, was released \cite{2019arXiv190201025L} and shown to be faster than Monte-Carlo based simulations for other diffusion sequences.

The discretization of the Bloch-Torrey equation by finite elements is a more recently developed approach 
for the purpose of dMRI simulations, in contrast to random walk simulations, which have a longer history.  
While finite element discretization is more difficult to implement than random walk simulations, 
the approach benefits from  from long-established theoretical and numerical developments 
by the mathematical and engineering communities.  
In particular, software packages for the automated solutions of partial differential equations using finite element discretization, such as FEniCS \cite{Logg2012, fenics:www}, are subject to active support and development. However, because diffusion MRI simulation is a relatively new application area, there is still a gap 
between the simulation needs of the MRI community and the available tools built on top of these finite elements software packages.  

The deployment of FEniCS containers \cite{FEniCSContainers} opens a new direction to improve productivity and sharing in the scientific computing community. In particular, it can dramatically improve the accessibility and usability of high-performance computing (HPC) systems. In this paper, we address two potential difficulties 
in using FEniCS for diffusion MRI simulation.  
First, we simplified software installation by the use of FEniCS containers that are completely portable across multiple platforms. 
Second, we provide a simulation framework written in Python and whose code is open source.  
This simulation framework can be seamlessly integrated with cloud computing resources such as 
Google Colaboratory notebooks (working on a web browser) or with Google Cloud Platform with MPI parallelization. 

One of the advantages of the simulation framework we propose here over the Matlab-based SpinDoctor \cite{2019arXiv190201025L} is that the Python code is free, whereas SpinDoctor requires the purchase of the 
software Matlab. Many researchers are now adopting Python since it is a free, cross-platform, general-purpose and high-level programming language. Plenty of Python scientific packages are available with extensive documentation such as SciPy for fundamentals of scientific computing, NumPy for large and multi-dimensional arrays and matrices, SymPy for symbolic computation, IPython for the enhanced interactive console, Pandas for data structures \& analysis, Matplotlib for comprehensive 2D plotting. In addition, parallel computing for finite elements is relatively easy to implement within FEniCS, thus, this framework has advantages over SpinDoctor for very large scale problems. 

The disadvantage of this simulation framework compared to SpinDoctor is the current lack of high-order adaptive time-stepping methods in Python tailored to the kind of ODEs systems coming from finite elements discretization, whereas such time-stepping methods are available in Matlab.
Thus, in contrast to SpinDoctor where an adaptive, variable order, time-stepping method is used, 
the time-stepping method in the proposed framework is the $\theta-$method, with a fixed time step size. The $\theta-$method can be second-order accurate if $\theta$ is chosen to be $\frac{1}{2}$.

The simulation framework we propose can meet the following simulation needs:
\begin{enumerate}
    \item the specification of an intrinsic diffusion tensor and a $T_2-$relaxation coefficient in each geometrical compartment;
    \item the specification of a permeability coefficient on the interface between the geometrical compartments;
    \item the periodic extension of the computational domain (assumed a box);
    \item the specification of general diffusion-encoding gradient pulse sequences;
    \item the simulation of thin-layer and thin-tube media using a discretization on manifolds;
\end{enumerate}
Since this framework is based on FEniCS, packaged as an image, it inherits all functionalities of FEniCS related to mesh generation, mesh adaptivity, finite elements matrices construction, linear system solve, 
solution post-processing and display, as well as the underlying FEniCS computational optimization related to the above tasks.
Finally, the framework is conceived with cloud computing and high performance computing in mind, thus, it
\begin{enumerate}
    \item supports Cloud Computing with Google Colaboratory and Google Cloud Platform;
    \item allows for MPI parallelization.
\end{enumerate}

The paper is organized as follows. In Section \ref{sec:theory} we recall the diffusion MRI simulation model based on the  Bloch-Torrey equation. Then, we propose a portable simulation framework in Section \ref{sec:method} for which the numerical validation and the comparison are carried out in Section \ref{sec:validation}. Several simulations examples are shown in Section \ref{sec:examples}. We share some perspectives about the proposed framework in Section \ref{sec:discussion_future_work}. The paper is finalised with a conclusion in Section \ref{sec:conclusion}.

\section{Theory}\label{sec:theory}
\begin{figure}[ht]
    \centering
    \subfloat[]{
        \begin{tikzpicture}
         \coordinate (c) at (-0.75,-0.75);
         \coordinate (d) at (0.75,0.75);
         \fill[draw, line width=0.3mm,top color=gray!30,bottom color=white] (2,2) rectangle (-2,-2);         
         
       	 \fill[draw=blue, line width=0.3mm,top color=gray!30,bottom color=gray, rounded corners=1mm] (c)           
         \irregularcircle{1cm}{1mm};
         \fill[draw=blue, line width=0.3mm,top color=gray!30,bottom color=gray, rounded corners=1mm] (d)
         \irregularcircle{0.9cm}{1mm};         
         \node[] at (c) {$\bm\Omega_0, \bm D_0$};      
		 \node[] at (d) {$\bm\Omega_0, \bm D_0$};      
		 \node[] at (1.2,-1.5) {$\bm\Omega_1, \bm D_1$};
		 \node[] at (0.8,-0.5) {\textcolor{blue}{$\bm\Gamma, \bm\kappa$}};

        \end{tikzpicture}
        \label{fig:composed_domain}
    }
    \subfloat[]{
        \includegraphics[width=0.6\textwidth]{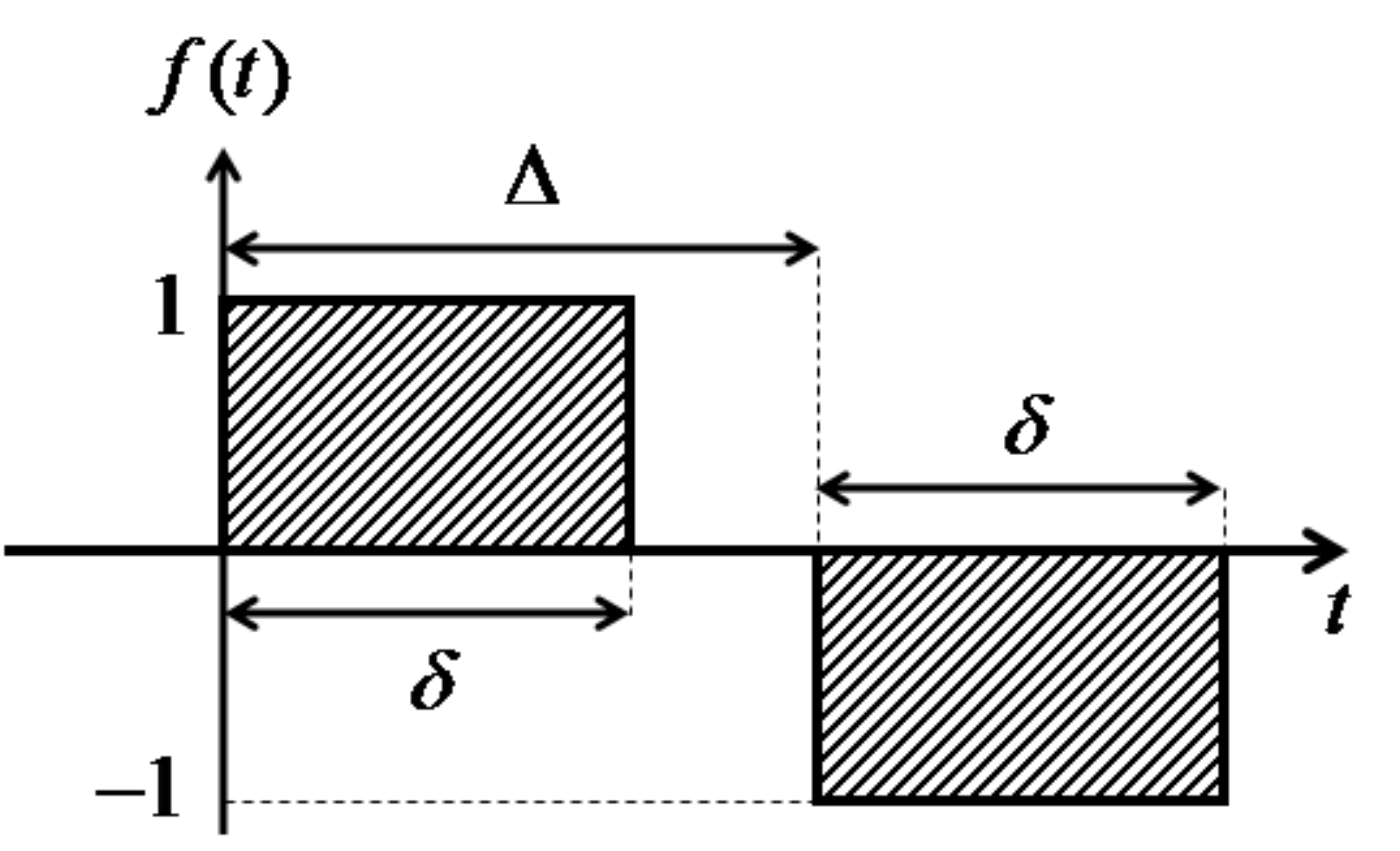}
        \label{fig:PGSE}
    }
    \caption{A composed domain $\Omega=\Omega_0\cup \Omega_1$ (a), and a PGSE sequence (b).}
    \label{fig:bio_model}
\end{figure}
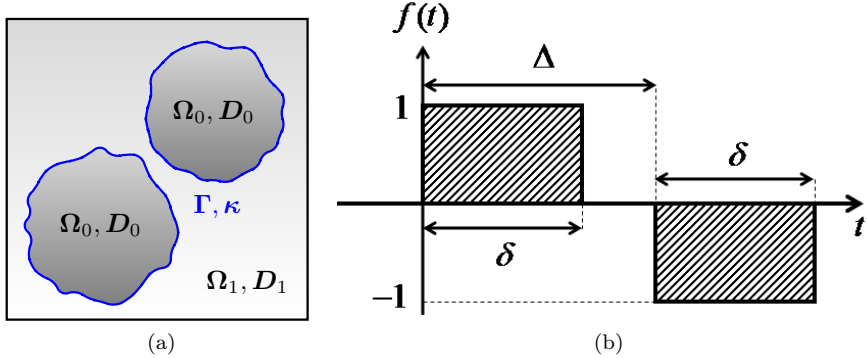

The evolution of the complex transverse magnetization $U(\bm x, t)$ over time $t$ can be described by the Bloch-Torrey equation \cite{PhysRev.104.563}. For simplicity we consider a medium composed of two compartments, $\Omega=\Omega_0\cup \Omega_1$, each of which may be disconnected (see Fig. \ref{fig:composed_domain}). The equation takes the following form 
\begin{equation}\label{eq:strong_BT}
\frac{\partial U(\bm x, t)}{\partial t}=-i\,\gamma f(t)\, \bm g \cdot \bm x \, U(\bm x, t)- \frac{U(\bx, t)}{T_2(\bx)} + \nabla \cdot \Bigl( \bm D(\bm x) \, \nabla U(\bm x, t)\Bigl)
\end{equation}
where $i$ is the complex unit ($i^2=-1$), $\gamma=2.67513\times 10^8\,\rm rad\,s^{-1}T^{-1}$ is the
gyromagnetic ratio of the water proton, and $\bm g$ is the diffusion gradient including gradient strength $g=\|\bm g\|$ and gradient direction $\bm q=\frac{\bm g}{\|\bm g\|}$.  
In the general case, ${\bm D}(\bm x)$ is the diffusion tensor, a symmetric positive definite $3 \times 3$ matrix.  
$T_2$ relaxation is the process by which the transverse magnetization decays or dephases.

On the interfaces between different compartments
the magnetization is allowed to be discontinuous
via the use of a permeability coefficient $\kappa$ \cite{doi:10.1063/1.436751}
\begin{equation}
\label{eq:btpde_ic}
\begin{aligned}
        \Bigl\llbracket \bm D \,\nabla U\cdot \bm n^0 \Bigl\rrbracket & = 0\\
        \Bigl\{ \bm D \,\nabla\, U\cdot \bm n^0 \Bigl\} & = -\kappa \,\llbracket U \rrbracket
\end{aligned}
\end{equation}
for $\bm x \in \Gamma=\partial\Omega_0\cap\partial\Omega_1$ and $\bm n^k$ is a normal vector pointing
outward $\Omega_k$.

The temporal profile $f(t)$ can vary for different applications and the most commonly used diffusion-encoding sequence in diffusion MRI literature is called the Pulsed-Gradient Spin Echo (PGSE) sequence \cite{Stejskal1965}.  For this sequence, one can write $f(t)$ in the following way (see also Fig. \ref{fig:PGSE}):
\begin{equation}\label{eq:pgse}
f(t) =
\begin{cases}
1, \quad &0 \leq t \leq \delta, \\
-1,
\quad & \Delta < t \leq \Delta+\delta,\\
0, \quad & \text{otherwise.}
\end{cases}
\end{equation}
The quantity $\delta$ is the duration of the diffusion-encoding gradient pulse and 
$\Delta$ is the time delay between the start of the two pulses.
Beyond the PGSE, the Oscillating Gradient Spin Echo (OGSE) \cite{Does2003}, nonstandard diffusion sequences such as double diffusion encoding \cite{Shemesh2016, Dhital2019, Novikov2019, Henriques2019} and multidimensional diffusion encoding \cite{TOPGAARD201798} can be modelled.

Concerning the boundary conditions (BCs) on the exterior boundaries $\partial \Omega$, 
there are two options that are very often employed.  
One is placing the spins to be simulated sufficiently away from $\partial \Omega$
and impose simple BCs on $\partial \Omega$ such as 
homogeneous Neumann conditions. 
This supposes that the spins would have a low probability of having arrived 
at $\partial \Omega$ during the diffusion experiment. Another option is to place the spins anywhere desired, but to assume that $\Omega$ is repeated periodically in all space directions to fill $\mathbb{R}^d$,
for example, $\Omega=\prod_{k=1}^d[a_k,b_k]$. So, one can mimic the phenomenon where the water molecules can enter and exit the computational domain. Under this assumption of periodic continuation of the geometry, the 
magnetization satisfies 
pseudo-periodic BCs on $\partial \Omega$ \cite{Xu2007}
\begin{equation}
\label{eq:btpde_bc}
\begin{aligned}
U_m&=U_s e^{{i\,\theta_k(t)}},\\ 
\bm D_m\,\nabla U_m\cdot \bm n
&=\bm D_s\,\nabla U_s\cdot \bm n\,e^{{i\,\theta_k(t)}},
\end{aligned}
\end{equation}
where 
$$
U_m=\left. U(\bm x,t)\right|_{x_k=a_k},\quad U_s=\left.U(\bm x,t)\right|_{x_k=b_k}
$$
$$
\nabla U_m\cdot \bm n
=\nabla U(\bm x,t)\cdot \bm n\,\biggl|_{x_k=a_k}, \quad
\nabla U_s\cdot \bm n=\nabla U(\bm x,t)\cdot \bm n\,\biggl|_{x_k=b_k}
$$
and
$$\theta_k(t) := \gamma \;g_k\,(b_k-a_k) \,\mathcal{F}(t), k=1,\cdots,d, \quad \mathcal{F}(t)=\int\limits_0^t f(s)\,ds.$$
Here we use `m' and `s' to indicate master and slave components of the pseudo-periodic BCs. The master-slave method corresponds to the implementation of the periodic BCs \cite{2008IJNME73361Y}.

The MRI signal $S$ is the total transverse magnetization $U(\bm x, t)$ over $\Omega$ 
measured at the echo time $T$ 
\begin{equation}\label{eq:signal}
S= \int_{\bm x\in \Omega} U(\bm x, T)\; d\bm x 
\end{equation}
The signal is usually plotted against the gradient strength $g=\|\mathbf{g}\|$ or a quantity called the $b$-value which is defined as the following
\begin{equation}\label{eq:b_value}
  b=\gamma^2\|\mathbf{g}\|^2\int\limits_0^T {\mathcal{F}(s)^2}\,ds.
\end{equation}

\section{Method}\label{sec:method}
For software portability, we consider two container technologies which are Docker \cite{BashariRad2017} and Singularity \cite{10.1371/journal.pone.0177459}.  They allow for bundling the whole collection of software packages that a user needs in a single file, that can be shared and used by collaborators. This would make a huge impact in scientific applications, where reproducibility is a core concern \cite{NAP25303}. In particular, this enables us to develop software that other users can easily test. A software update reduces to a matter of downloading the newest version of a single file and different versions can coexist next to each other for easy consistency checks. We choose Docker for the IPython notebooks and Singularity for the deployment on HPC infrastructure. They follow the same workflow as the following
\begin{center}
\begin{tikzpicture}[node distance=1.6cm]
\node (box1) [startstop] { Setting the working environment };
\node (box2) [startstop1, below of=box1] {Pre-processing the meshes and other input parameters};
\node (box3) [startstop, below of=box2] {Solving the Bloch-Torrey equation};
\node (box4) [startstop, below of=box3] {Post-processing};
\draw [arrow] (box1) -- (box2);
\draw [arrow] (box2) -- (box3);
\draw [arrow] (box3) -- (box4);
\end{tikzpicture}    
\end{center}

\subsection{Diffusion MRI simulation library}
The solution of the Bloch-Torrey equation and other functionalities related to diffusion MRI simulations have been packaged into Python library \lstinline{DmriFemLib}, saved in GitHub.
\begin{center}
    \small{\url{https://github.com/van-dang/DMRI-FEM-Cloud/blob/master/DmriFemLib.py}}
\end{center}
Due to considerations related to the way FEniCS envisions the PDE solution workflow, and the fact that the PDE from the diffusion
MRI simulation problem has some important differences from the typical PDEs for which FEniCS was designed, we made the following choices regarding the implementation of the numerical method that are different than the choices made in the Matlab-based toolbox SpinDoctor. These choices are:
\begin{enumerate}
    \item the permeable interface conditions are imposed by the use of the partition of unity finite element method (PUFEM) \cite{MELENK1996289, NGUYEN2018271};
    \item the pseudo-periodic BCs coming from the periodic extension of the computational box are imposed on either side of the box face by a PDE transformation;
    \item in case of a non-periodic mesh, the necessary pseudo-periodic BCs are imposed by using an artificially permeability coefficient on the box face whose magnitude is inversely proportional to the finite element mesh size \cite{Nguyen1080573, NGUYEN2018271};
    \item the implicit Crank-Nicolson method is chosen as the time-stepping method. It is especially important to ensure the stability with the use of the artificial permeability coefficient.
\end{enumerate}
\subsection{Mesh generation}
Dealing with meshes is one of the most challenging problems in FEM and we inherit what has been done in Python and FEniCS regarding this issue. For simple geometries, one can internally use some built-in meshes. Meshing a box $\Omega=[0,10]^3$ with given resolutions \lstinline{nx, ny, nz} is simply done by the following commands
\begin{lstlisting}
nx, ny, nz = 10
mesh = BoxMesh(Point(0.0, 0.0, 0.0), Point(10.0, 10.0, 10.0), nx, ny, nz)
\end{lstlisting}
For more complicated geometries, it is recommended to use \lstinline{mshr} \cite{BenjaminKehlet}, the mesh generation component of FEniCS, to generate simplicial DOLFIN meshes in 2D and 3D from geometries described by Constructive Solid Geometry or from surface files, utilizing CGAL and Tetgen as mesh generation backends. The commands below are used to generate a two-layered disk:
\begin{lstlisting}
from mshr import *
R1, R2 = 5, 10; origin = Point(0.,0.);
circle = Circle(origin, R1, segments=32)
domain = Circle(origin, R2, segments=32)
domain.set_subdomain(1, circle)
mesh = generate_mesh(domain, 15) # 15 is the resolution
\end{lstlisting}
More generally, our framework accepts meshes in the DOLFIN XML format \cite{Logg:2010:DAF:1731022.1731030}. In this paper, the meshes were generated with GMSH \cite{geuzaine09gmsh}, Salom\'e \cite{salome}, and ANSA \cite{ansa}. The GMSH script (\lstinline{.geo}) and Salom\'e script (\lstinline{.py}) are available at
\begin{center}
\small{\url{https://github.com/van-dang/DMRI-FEM-Cloud/tree/mesh}}
\end{center}
and they are distributed through examples discussed later in the paper. GMSH can be embedded in our framework
\begin{lstlisting}
import os
# define mesh_name ...
os.system('gmsh -3 '+mesh_name+'.geo -o '+mesh_name+'.msh')
\end{lstlisting}

All the formats need to be converted to DOLFIN XML format by the use of either \lstinline{dolfin-convert} available with FEniCS or MESHIO \cite{meshio}. To convert a mesh from \lstinline{.msh} to \lstinline{.xml} in a Colaboratory notebook, we just simply call
\begin{lstlisting}
import os
os.system("dolfin-convert mesh.msh mesh.xml")
\end{lstlisting}

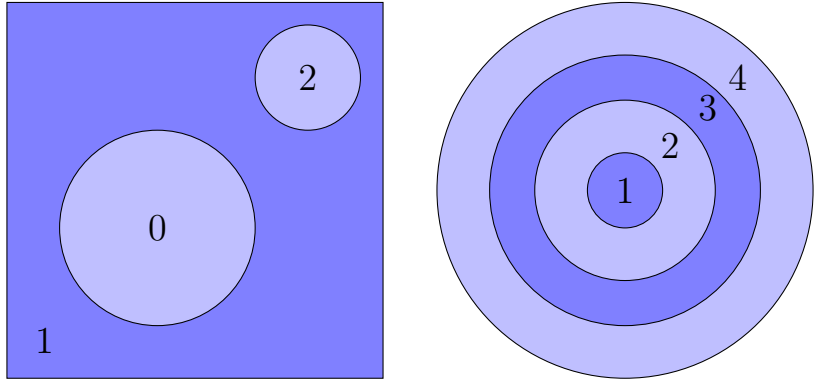
\begin{figure}[ht]
\centering
\begin{tikzpicture}
  \draw [fill=MyColorOne] (0,0) rectangle ++(5,5);
  \node[] at (0.5,0.5) {{\Large{1}}};
  \draw[fill=MyColorOneLight] (2,2.) circle [radius=1.3] node {\Large{0}};
  \draw[fill=MyColorOneLight] (4.0,4.0) circle [radius=0.7] node {\Large{2}};
\end{tikzpicture}\quad\quad
\begin{tikzpicture}
\draw[fill=MyColorOneLight] (0.0,0.0) circle [radius=2.5];
\node[] at (1.5,1.5) {{\Large{4}}};
\draw[fill=MyColorOne] (0.0,0.0) circle [radius=1.8];
\node[] at (1.1,1.1) {{\Large{3}}};
\draw[fill=MyColorOneLight] (0.0,0.0) circle [radius=1.2];
\node[] at (0.6,0.6) {{\Large{2}}};
\draw[fill=MyColorOne] (0.0,0.0) circle [radius=0.5] node {\Large{1}};
\end{tikzpicture}
\caption{For multi-compartment domains, the compartments need to be sorted into two groups, \lstinline{oddgroup} (blue) marked with odd numbers and \lstinline{evengroup} (light-blue) marked with even numbers which are referred to as the \lstinline{partition_marker} such that in each group, the compartments should be completely disconnected. It is, therefore, enough to use a \lstinline{phase} function with two values 0 and 1 to impose the permeability.}
\label{fig:partition_marker}
\end{figure}

For a multi-compartment domain, a \lstinline{partition_marker} is used to assign each compartment to an identity that allows for defining nonuniform initial conditions, discontinuous diffusion tensors and discontinuous $T_2-$relaxation.
It is a \lstinline{MeshFunction} in FEniCS defined as the following 
\begin{lstlisting}
partition_marker = MeshFunction("size_t", mesh, mesh.topology().dim())
for cell in cells(mesh):
    partition_marker[cell.index()] = <an identity>;
\end{lstlisting} 
To impose the interface conditions between compartments, we use a \lstinline{phase} function which is $\Phi_h$ in Eq. (\ref{eq:phase_func}). This function initially supports two-compartment domains since it has only two values 0 and 1. To apply for a multi-compartment domain, the compartments need to be sorted into two groups \lstinline{oddgroup} and \lstinline{evengroup}. The permeability is imposed at the intermediate interfaces between the two groups. In each group, however, there is no interface between two compartments or in other words, they are completely disconnected (see Fig. \ref{fig:partition_marker}).

We defined a routine called \lstinline{CreatePhaseFunc} to create the \lstinline{phase} function and the \lstinline{partition_marker}. If the sub-meshes corresponding to compartments are given, these two functions can be created as follows
\begin{lstlisting}
# Download the meshes
mesh = Mesh("multi_layer_torus.xml")  
cmpt_mesh = Mesh("multi_layer_torus_compt1.xml")
evengroup = []
oddgroup = [cmpt_mesh]
phase, partion_list, partition_marker = CreatePhaseFunc(mesh, evengroup, oddgroup, None)                
\end{lstlisting}
The  \lstinline{partition_marker} can be generated and saved to a DOLFIN XML file, for instance \lstinline{partition_marker.xml}. The file structure below shows that the elements (cells) with indices of 0 and 1 are assigned with partition marker 3 and 4 respectively.
\begin{lstlisting}
<?xml version="1.0"?>
<dolfin xmlns:dolfin="http://fenicsproject.org">
  <mesh_function>
    <mesh_value_collection type="uint" dim="3" size="6614">
      <value cell_index="0" local_entity="0" value="3" />      
      <value cell_index="1" local_entity="0" value="4" />
      ...
    </mesh_value_collection>
  </mesh_function>
</dolfin>
\end{lstlisting}
In case the partition markers are defined with GMSH by the use of ``physical groups", the file can be simply generated by calling our built-in routine \lstinline{GetPartitionMarkers}
\begin{lstlisting}
GetPartitionMarkers("mesh.msh", "partition_marker.xml")
\end{lstlisting}
With a given \lstinline{partition_marker.xml}, the phase function is generated by
\begin{lstlisting}
File("partition_marker.xml")>>partition_marker
phase, partition_list = CreatePhaseFunc(mesh, [], [] partition_marker)
\end{lstlisting}
 It requires extra care to generate periodic meshes to use the strong imposition of the pseudo-periodic BCs (see \ref{sec:strong_bc_fem}).  GMSH supports this by \lstinline{Periodic} mapping which is equivalent to \lstinline{Projection} routine in Salom\'e. As part of the framework, we developed the scripts to generate periodic meshes with cells are available in GMSH
 \begin{center}
     {\small\url{https://github.com/van-dang/DMRI-FEM-Cloud/blob/mesh/CirclesInSquare.geo}}
\end{center}
and in Salom\'e
 \begin{center}
     {\small\url{https://github.com/van-dang/DMRI-FEM-Cloud/blob/mesh/SpheresInBox.py}}.
 \end{center}
 
\subsection{The main workspace}
The workflow is carried out in the main workspace which is either web-based Jupyter notebooks or a script-based interface. Library \lstinline{DmriFemLib} and other functionalities need to be loaded here 
\begin{lstlisting}
import os
os.system("wget https://raw.githubusercontent.com/van-dang/ DMRI-FEM-Cloud/master/DmriFemLib.py")
from DmriFemLib import *
\end{lstlisting}
\subsubsection*{Python notebooks}
Google Colaboratory \cite{colab} is a free Jupyter notebook environment that requires no setup and runs entirely in the cloud. It can connect to either a hosted runtime provided by Google Cloud or a local runtime. The hosted runtime allows us to access free resources for up to 12 hours at a time and the current one, used to obtain the results presented in this paper, has the following configuration: 
\begin{itemize}
	\item Operating system: Ubuntu 18.04.2 LTS
	\item Processors: 2 x Intel(R) Xeon(R) CPU @ 2.30GHz
	\item RAM: 13 GB
\end{itemize}
In order to check the configuration you can run the following commands.
\begin{lstlisting}
!cat /proc/meminfo #check RAM memory
!lscpu #check processor
!cat /etc/lsb-release #check distribution
\end{lstlisting}

\begin{figure}[ht]
    \centering
    \includegraphics[width=0.6\textwidth]{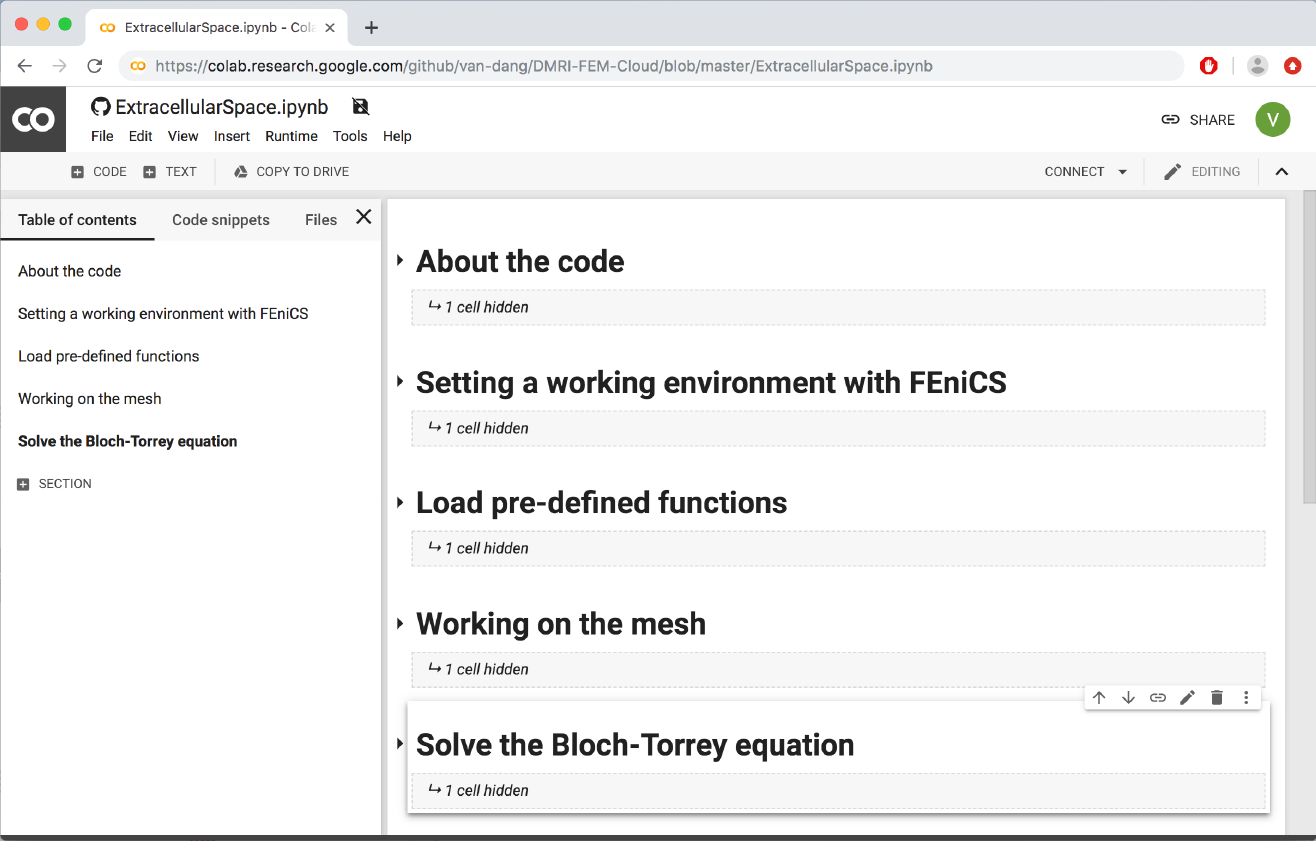}
    \caption{A typical Google Colaboratory notebook for diffusion MRI simulation.}
    \label{fig:colab}
\end{figure}

 Fig. \ref{fig:colab} shows a typical structure of our Google Colaboratory notebooks where the simulations can run directly since the setup of the FEniCS environment is done within the notebook.

The installation of FEniCS is quite straightforward in the hosted runtime. The command lines are just the same as the installation on Ubuntu.
\begin{lstlisting}
!apt-get install software-properties-common
!add-apt-repository ppa:fenics-packages/fenics
!sudo apt-get update
!apt-get install --no-install-recommends fenics
from dolfin import *; from mshr import *
\end{lstlisting}
For longer executions, it is more convenient to connect to the local runtime.
To this end, one can execute the following command lines to create a local runtime to which the notebook can connect. This command creates a Docker container from the latest stable FEniCS version at the time of writing, given with the \lstinline{fenics_tag} variable. Inside this container, we install a Jupyter extension developed by Google Colaboratory's developers and then run a Jupyter notebook from within the container on port \lstinline{8888}.

\begin{lstlisting}
fenics_tag=2019.1.0.r3 # version of FEniCS image
docker run --name notebook-local -w /home/fenics -v $(pwd):/home/fenics/shared -ti -d -p 127.0.0.1:8888:8888 quay.io/fenicsproject/stable:${fenics_tag} "sudo pip install jupyter_http_over_ws; sudo apt-get install -y gmsh; jupyter serverextension enable --py jupyter_http_over_ws; jupyter-notebook --ip=0.0.0.0 --NotebookApp.allow_origin='https://colab.research.google.com' --NotebookApp.port_retries=0 --NotebookApp.allow_root=True --NotebookApp.disable_check_xsrf=True --NotebookApp.token='' --NotebookApp.password='' --port=8888"
\end{lstlisting}

\subsubsection*{Script-based interface}
For the script-based interface with parallel executions, the workspace is available at
\begin{center}
     \small{\url{https://github.com/van-dang/DMRI-FEM-Cloud/blob/master/GCloudDmriSolver.py}}
 \end{center}
Users can pre-process the inputs for one- and multi-compartment domain by respectively using the functions implemented at
\begin{center}
{\small\url{https://github.com/van-dang/DMRI-FEM-Cloud/blob/master/PreprocessingOneCompt.py}}
\\
{\small\url{https://github.com/van-dang/DMRI-FEM-Cloud/blob/master/PreprocessingMultiCompt.py}}
\end{center}
This workspace can work with Docker image by launching the following command from a Mac or Linux terminal
\begin{lstlisting}
docker run -ti -p 127.0.0.1:8000:8000 -v $(pwd):/home/fenics/shared -w /home/fenics/shared quay.io/fenicsproject/stable:current
\end{lstlisting}
However, in the HPC context, Singularity is preferable to Docker due to the security, the accessibility, the portability, and the scheduling issues \cite{10.1371/journal.pone.0177459}. Fortunately, it is straightforward to build a Singularity image from a Docker image and for our framework, the command lines are as follows
\begin{lstlisting}
wget https://raw.githubusercontent.com/van-dang/DMRI-FEM-Cloud/ singularity_images/Singularity_recipe_FEniCS_DMRI
sudo singularity build -w writable_fenics_dmri.simg Singularity_recipe_FEniCS_DMRI
\end{lstlisting}

\subsubsection*{Code structure}
Although the interfaces are different between the web-based and the script-based workspaces, they have similar structures with three main classes
\begin{itemize}
    \item \lstinline{MRI_parameters} manages the diffusion pulses such as sequence type, $b-$values, $g-$value and the conversion between them.
    \item \lstinline{MRI_domain} manages the finite element meshes, function spaces, domain sizes, diffusion tensors, permeability, and boundary markers.
    \item \lstinline{MRI_simulation} manages the initial conditions, time-stepping sizes, linear solvers, solutions, and post-processing.
\end{itemize}
In \lstinline{MRI_domain}, the boolean variable \lstinline{IsDomainMultiple} is used to switch between the single-compartment and the multi-compartment domains. Both strong and weak imposition of the pseudo-periodic BCs have some advantages and disadvantages. The strong imposition works efficiently with periodic meshes with higher accuracy compared to the weak imposition. In some cases, it is, however, not practical to generate periodic meshes. We allow for both options by the use of a boolean variable \lstinline{IsDomainPeriodic}. When it is \lstinline{True}, Eq. (\ref{eqn:trBlo}) is solved, otherwise, Eq. (\ref{eq:strong_BT}) is solved.

In what follows, we show how to define an arbitrary diffusion sequence and how to use \lstinline{partition_marker} to define some input parameters on heterogeneous domains.

\subsubsection*{General diffusion-encoding sequence}

The framework allows for arbitrary temporal profiles $f(t)$. During the simulation, we need to compute its integral $\displaystyle \mathcal{F}(t)=\int_0^t f(s)\,ds$ and convert between $b-$value and the gradient strength $g-$value following Eq. (\ref{eq:b_value}). In the Python version, \lstinline{SymPy} is used to compute the symbolic integration. So, users only need to provide the expression of $f(s)$ to the function member \lstinline{fs_sym} of the class \lstinline{MRI_parameters()}. $\mathcal{F}(t)$ and the conversion between $b-$value, $g-$value are automatically done. For example, a cos-OGSE sequence
\begin{equation}
    f(s)=\begin{cases}
    \cos(\omega\, s), & \mbox{ if } s\leq \delta\\
    -\cos\Bigl(\omega (s-\tau)\Bigl), & \mbox{ if }  \tau < s \leq \delta+\tau\\
    0, & \mbox{otherwise}
    \end{cases}
\end{equation}
with $\displaystyle\omega=\frac{2\,n\pi}{\delta}, \tau=\frac{\delta+\Delta}{2}$ can be simply defined as the following
\begin{lstlisting}
import sympy as sp
mp = MRI_parameters()
...
mp.delta, mp.Delta = 10000, 10000 
mp.T = mp.delta+mp.Delta
mri_para.nperiod = 2
omega = 2.0*mri_para.nperiod*pi/mri_para.delta
tau = mp.T/2.
mri_para.fs_sym = sp.Piecewise(
    (  sp.cos(omega*mri_para.s) ,       mri_para.s <= mri_para.delta ),
    (  0.,                              mri_para.s <= tau ),
    (  -sp.cos(omega*(mri_para.s-tau)), mri_para.s <= mri_para.delta + tau ),
    (  0., True )  
)
...
mp.Apply() # F(t) and the conversion between b and q are done here
\end{lstlisting}

\subsubsection*{Initial conditions}\label{sec:initial_conditions}
By default, we initialize the spins to be one everywhere. However, it is possible to define discontinuous initial conditions which are illustrated in the following code snippet.
\begin{lstlisting}
mri_simu = MRI_simulation()
mri_para = MRI_parameters()
mymesh = Mesh(...)
mri_domain = MRI_domain(mymesh, mri_para)
...
IC_array = [0, 1, 0];
dofmap_DG = mri_domain.V_DG.dofmap()
disc_ic = Function(mri_domain.V_DG);
for cell in cells(mymesh):
    cmk = partition_marker[cell.index()]
    cell_dof = dofmap_DG.cell_dofs(cell.index())
    disc_ic.vector()[cell_dof] = IC_array[cmk]; 
disc_ic=project(disc_ic, mri_domain.V)
...
mri_simu.solve(mri_domain, mri_para, linsolver, disc_ic)
\end{lstlisting}

\subsubsection*{Diffusion coefficients and tensors}\label{sec:diffusion_tensors}
We allow for a general definition of the diffusion tensor  with $d\times d$ components
\begin{equation}
    \bm D(\bx)=\Bigl[ d_{jk}(\bx) \Bigl]_{j=1..d, k=1..d}
\end{equation}
where $d_{jk}(\bx)$ is cell-based piecewise continuous. We loop through all elements (cells) and the value can be determined by the coordinates of the cell midpoint \lstinline{p=cell.midpoint()} or a given \lstinline{partition_marker}.

\begin{lstlisting}
partition_marker = MeshFunction("size_t", mesh, mesh.topology().dim())
# define partition markers
...
D0_array=[3e-3, 1e-3, 3e-3]
# Variable diffusion tensor
V_DG=mri_domain.V_DG; dofmap_DG = V_DG.dofmap(); 
d00 = Function(V_DG); d01 = Function(V_DG); d02 = Function(V_DG)
d10 = Function(V_DG); d11 = Function(V_DG); d12 = Function(V_DG)
d20 = Function(V_DG); d21 = Function(V_DG); d22 = Function(V_DG)
for cell in cells(mymesh):
    p = cell.midpoint() # the coordinate of the cell center.
    cmk = partition_marker[cell.index()]
    cell_dof = dofmap_DG.cell_dofs(cell.index())
    d00.vector()[cell_dof] = D0_array[cmk]; 
    d11.vector()[cell_dof] = D0_array[cmk]; 
    d22.vector()[cell_dof] = D0_array[cmk];
mri_domain.ImposeDiffusionTensor(d00,d01,d02,d10,d11,d12,d20,d21,d22)
\end{lstlisting}
\subsubsection*{$T_2-$relaxation coefficient}
By default $T_2-$relaxation is set to be \lstinline{1e16}. However, users can define it similarly to the diffusion entry. The following code shows how to define $T_2$ for a three-compartment domain.
\begin{lstlisting}
T2_array=[4e16, 4e4, 4e4]
dofmap_DG = mri_domain.V_DG.dofmap()
T2 = Function(mri_domain.V_DG); 
for cell in cells(mymesh):
    cmk = partition_marker[cell.index()]
    cell_dof = dofmap_DG.cell_dofs(cell.index())
    T2.vector()[cell_dof] = T2_array[cmk]; 
mri_para.T2 = T2
\end{lstlisting}

\subsection{Solution visualization and post-processing}
After solving the Bloch-Torrey equation, the solutions are saved, visualized and the signals are computed following Eq. (\ref{eq:signal}) in a routine called \lstinline{PostProcessing}. Matplotlib \cite{matplotlib} is used for simple visualizations. For more advanced features, Paraview \cite{paraview} can be externally used for the saved solutions.

\section{Numerical validation and comparison}\label{sec:validation}
Unless stated otherwise, the simulations were performed for a PGSE with $\Delta=43100\tunit, \delta=10600\tunit$, $b-$values between 0 and $10000\bunit$, and the diffusion coefficient of $D=3\times10^{-3}\dunit$. The membrane between the compartments, if any, is permeable with the permeability of $\kappa=10^{-5}\kunit$.  The simulated signals are compared to the reference ones computed by the matrix formalism (MF) method \cite{GREBENKOV2010181}. 

We provide a complete simulation method of diffusion inside the multilayered structures such as concentric disks, cylinders, spheres, and torus with the mesh generation software GMSH \cite{geuzaine09gmsh}.

\begin{figure}[ht]
  \centering
     \subfloat[Initial conditions \text{[}0, 1, 0\text{]} \label{fig:discontinuous_ic010}]{\includegraphics[width=0.45\textwidth]{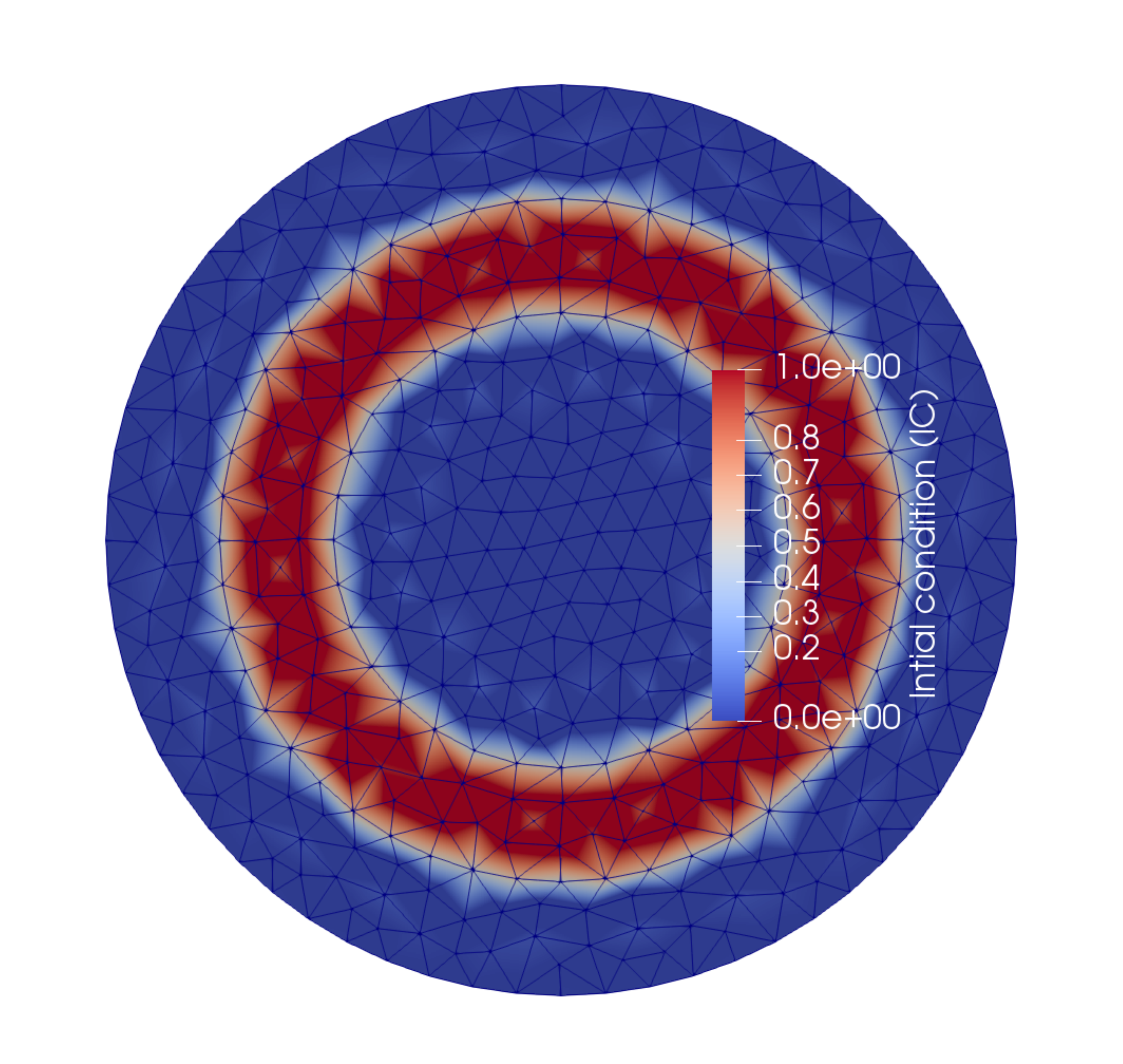}}
    \subfloat[Signals \label{fig:log_discontinuous_ic}]{\includegraphics[width=0.5\textwidth]{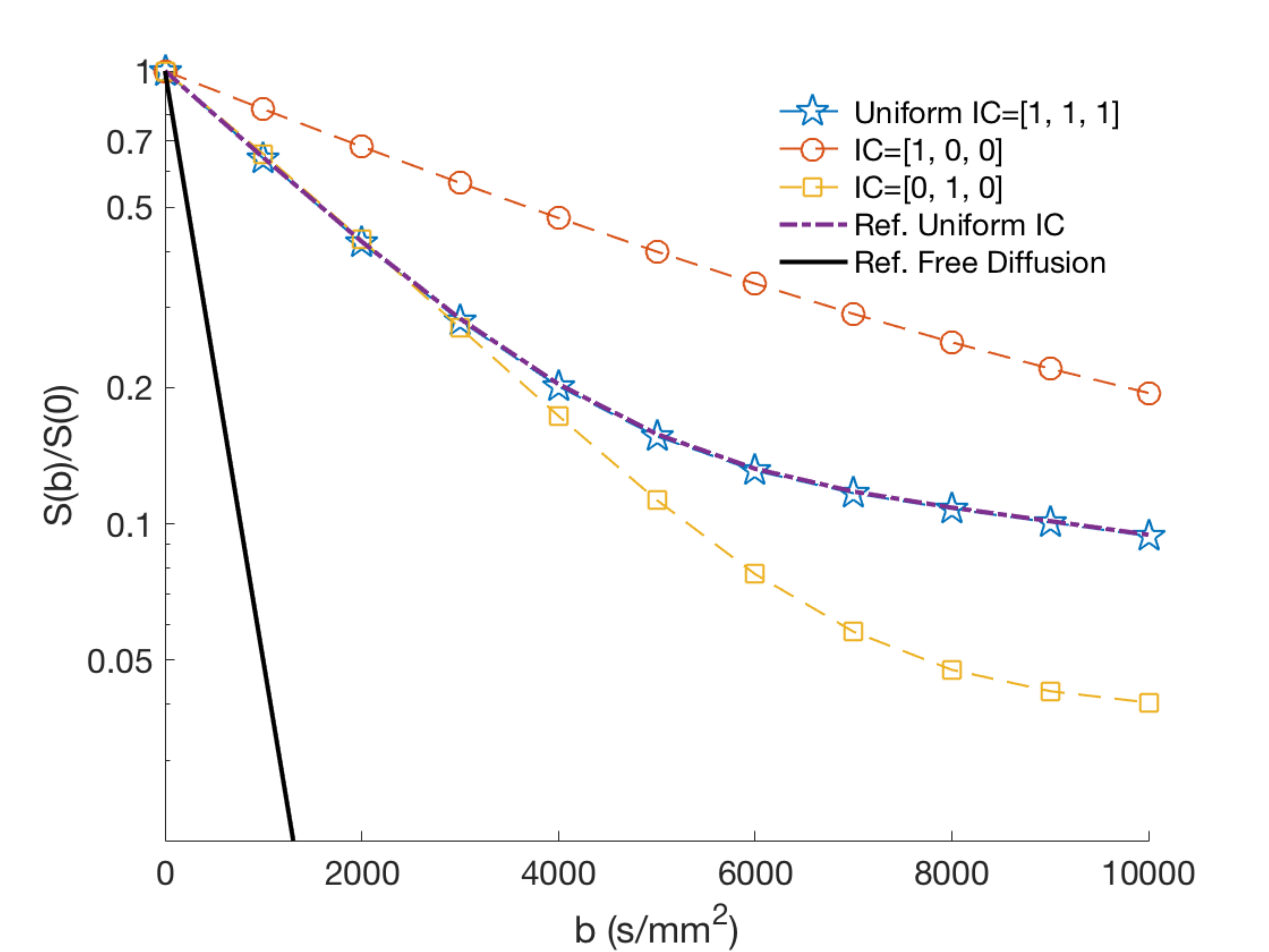}}
  \caption{Simulations of diffusion inside a three-layered disk of $[5, 7.5, 10]\lunit$ with different settings of the initial conditions for a PGSE sequence with $\Delta=43100\tunit$ and $\delta=10600\tunit$. Fig (a) show the initial conditions with \text{[}0, 1, 0\text{]}. The signals are strongly dependent on how the initial conditions are set up (b). The accuracy of the simulations is verified by comparing with the reference signal for the uniform distribution of the initial conditions.}
\label{fig:disc_initial_condition}
\end{figure}

First, we study diffusion inside a three-layered disk of $[5, 7.5, 10]\lunit$ with different settings of the initial conditions. Fig. \ref{fig:discontinuous_ic010} shows the setting \lstinline{IC_array=[0, 1, 0]} as discussed in Section \ref{sec:initial_conditions}.
The solver is available at
\begin{center}
    \small{\url{https://colab.research.google.com/github/van-dang/DMRI-FEM-Cloud/blob/master/DiscontinuousInitialCondition.ipynb}}
\end{center}
The time step size of $\Delta t=200\tunit$ is used. The signals are shown in Fig. \ref{fig:log_discontinuous_ic} where we show that the signals are strongly dependent on how we set up the initial conditions. The accuracy of the simulations is verified by comparing with the reference signal for the uniform distribution of the initial conditions.

\begin{figure}[ht]
    \centering
    \subfloat[A square box with circular cells\label{fig:periodic_rec_disks}]{
        \includegraphics[width=0.4\textwidth]{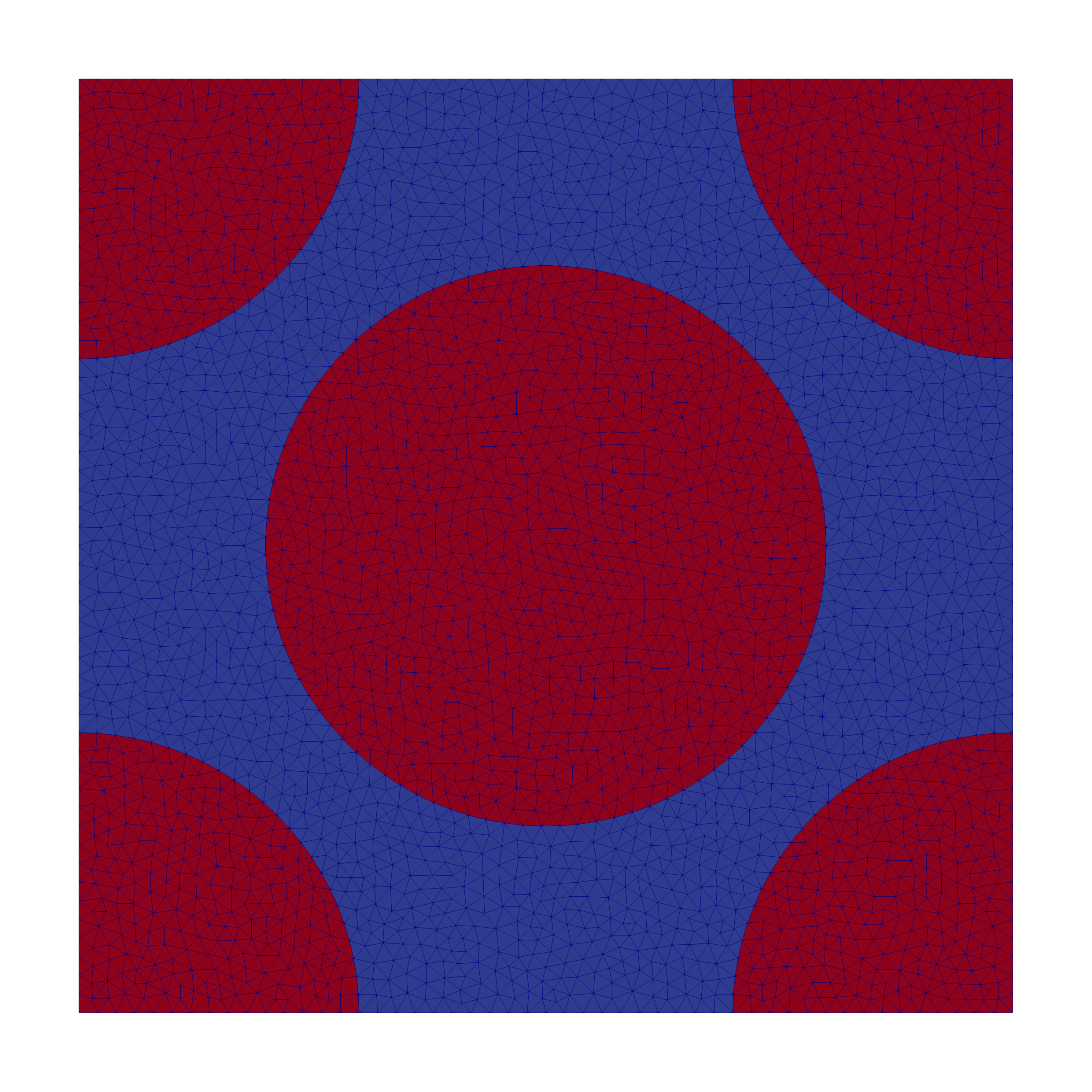}
    }
    \subfloat[Normalized signals \label{fig:signals_periodic_rec_disks}]{
        \includegraphics[width=0.5\textwidth]{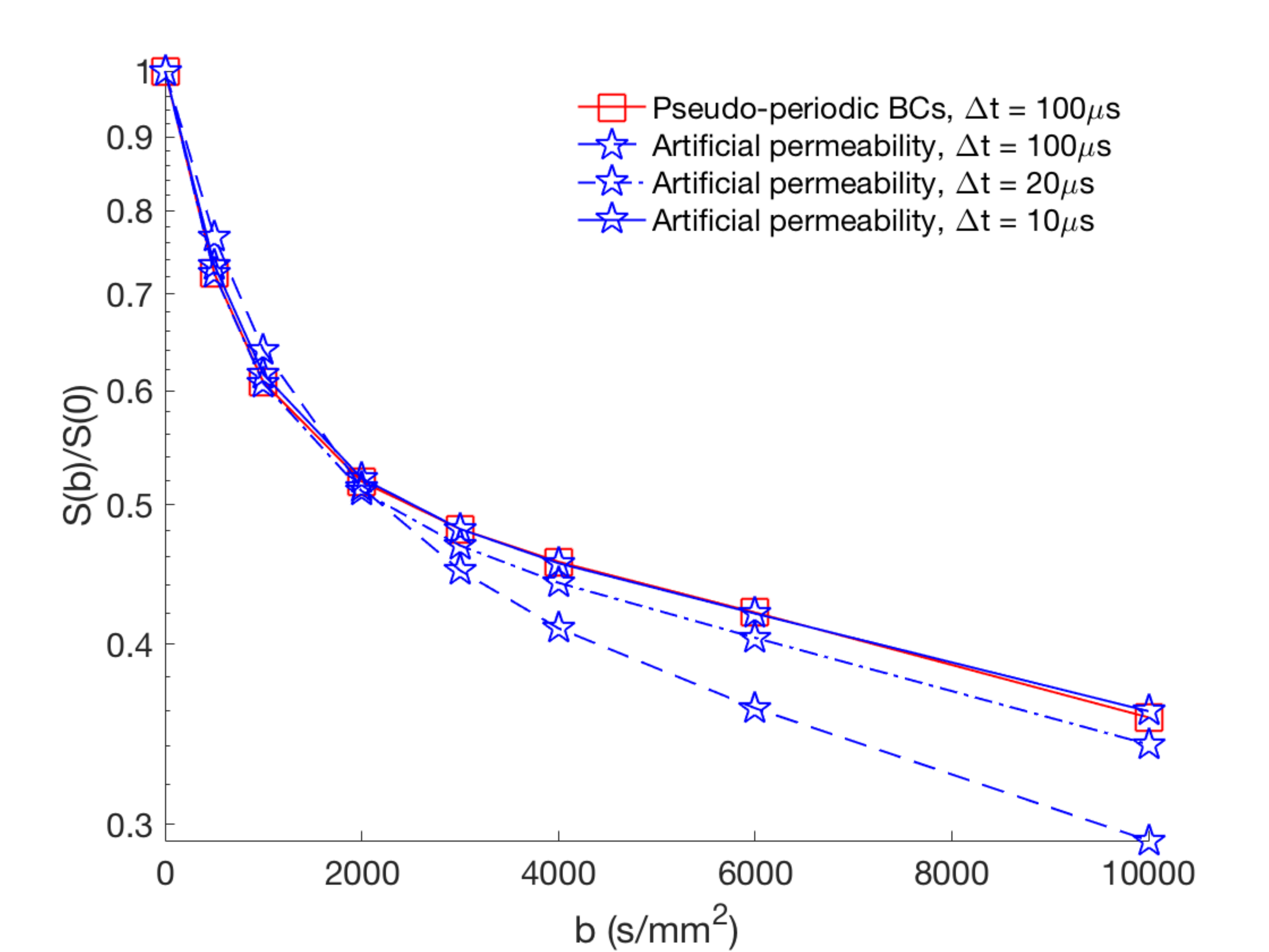}
    }    
    \caption{The artificial permeable method approaches the pseudo-periodic BCs. The time-step size needs to be small to achieve the same accuracy. It is, however, important to recall that the artificial permeable method is useful for non-periodic computational boxes.}
\end{figure}

To reduce the computational domain, one can assume that the domain is periodically repeated and our framework allows for imposing pseudo-periodic BCs. To illustrate this capacity, we consider a square $\Omega=[-5\lunit,5\lunit]^2$ including some permeable periodic cells with the permeability $\kappa=10^{-5}\kunit$ (see Fig. \ref{fig:periodic_rec_disks}). The signals were computed for a PGSE with $\Delta=13000\tunit, \delta=10000\tunit, {\bm q}=\frac{[1,1, 0]}{\sqrt{2}}$ (see Fig. \ref{fig:signals_periodic_rec_disks}. The solver is available at
\begin{center}
    \small{\url{https://colab.research.google.com/github/van-dang/DMRI-FEM-Cloud/blob/master/PeriodicDomains.ipynb}}
\end{center}
We see that the artificial-permeability method approaches the pseudo-periodic BCs. To achieve the same accuracy, the latter only needs $\Delta t=100\tunit$ which is  ten times as larger as the first. So, it is strongly recommended to use Eq. (\ref{eqn:trBlo}) if the domain is periodic. It is, however, important to recall that the first is useful for non-periodic computational boxes.

\begin{figure}[ht]
    \centering
    \subfloat[Three-layered disk]{
        \includegraphics[width=0.45\textwidth]{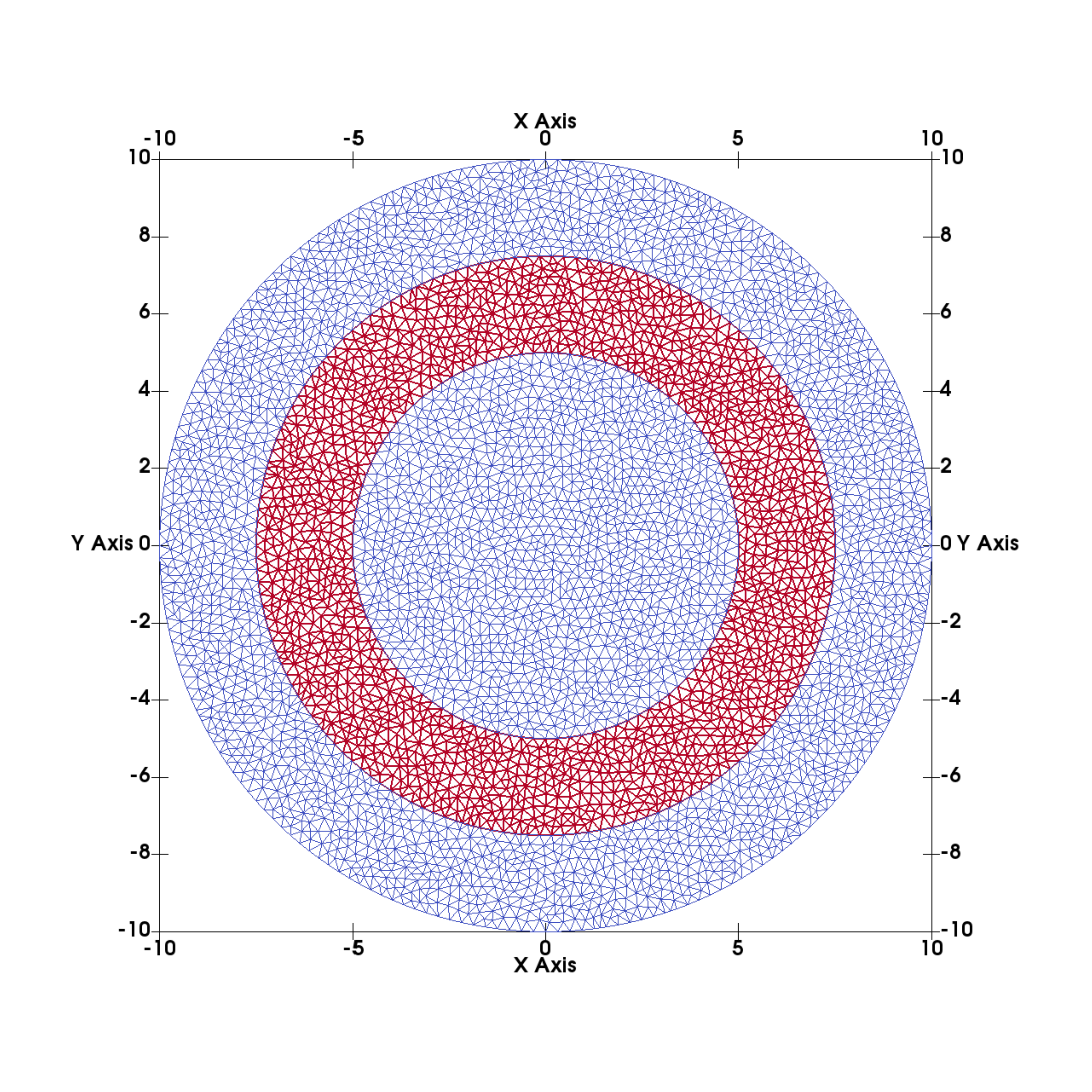}
        \label{fig:circle_3layers}
    }
    \subfloat[Three-layered sphere]{
        \includegraphics[width=0.5\textwidth]{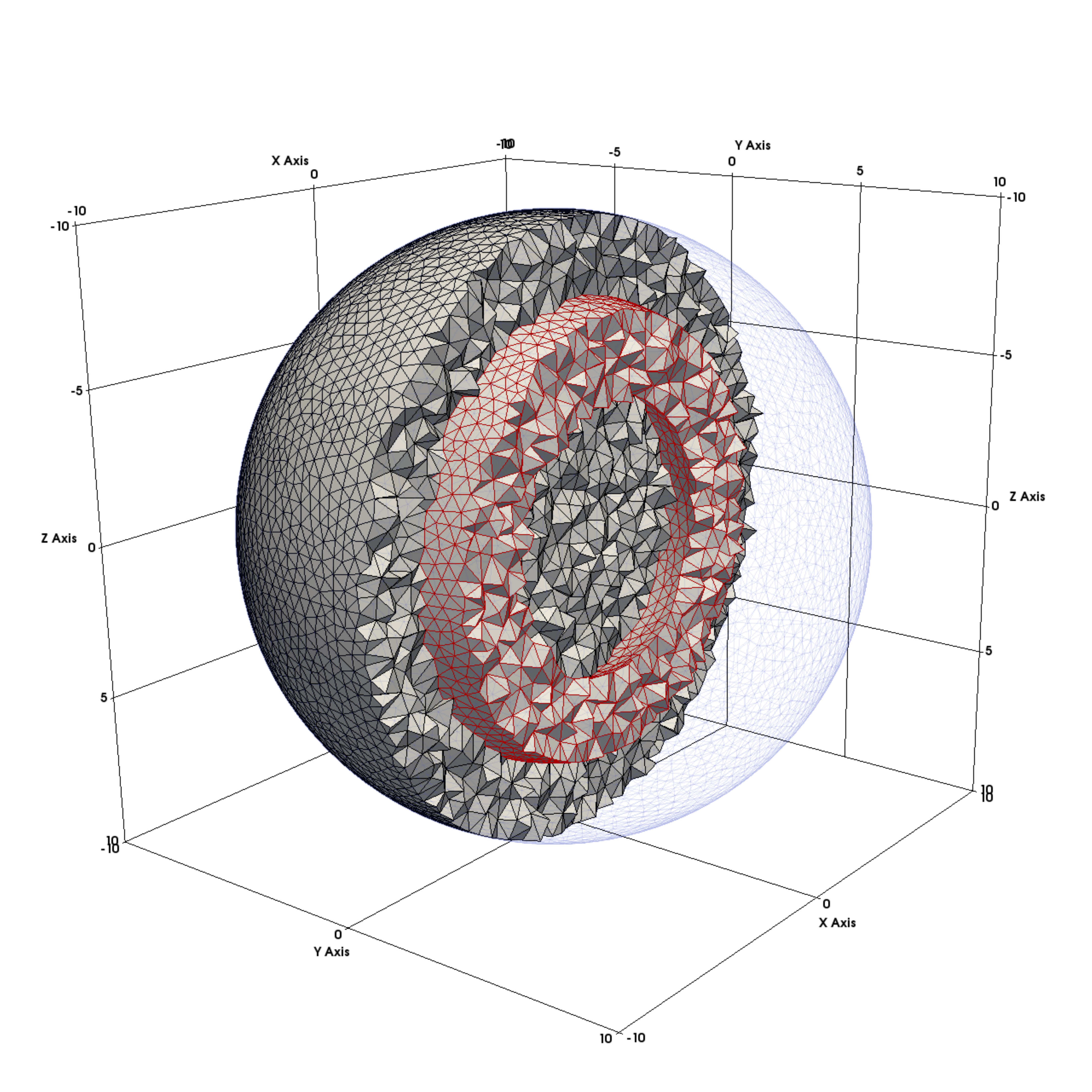}
        \label{fig:sphere_3layers}
    } \\      
    \subfloat[Three-layered torus]{
        \includegraphics[width=0.4\textwidth]{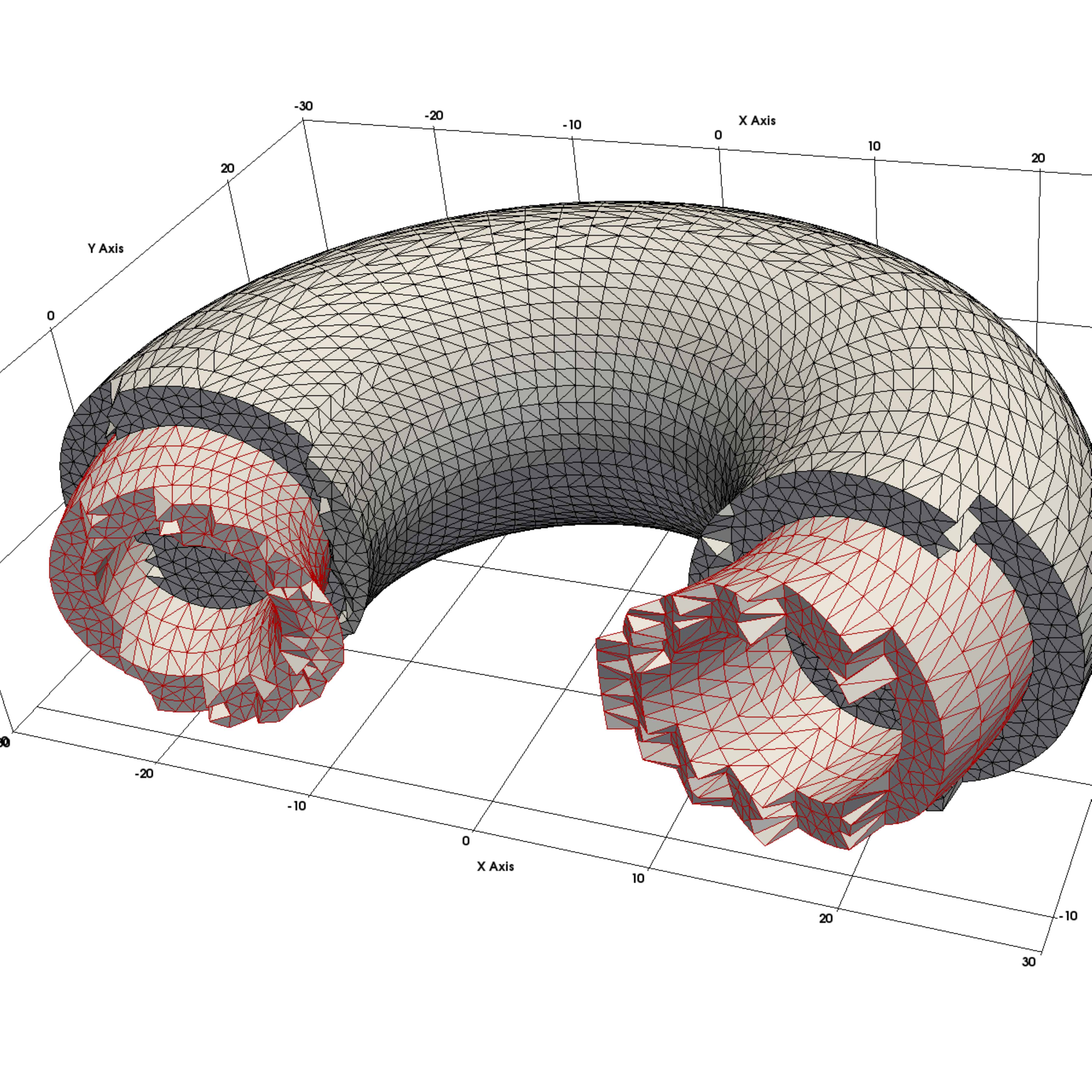}
        \label{fig:theta_torus}
    }
    \subfloat[Signals]{
        \includegraphics[width=0.5\textwidth]{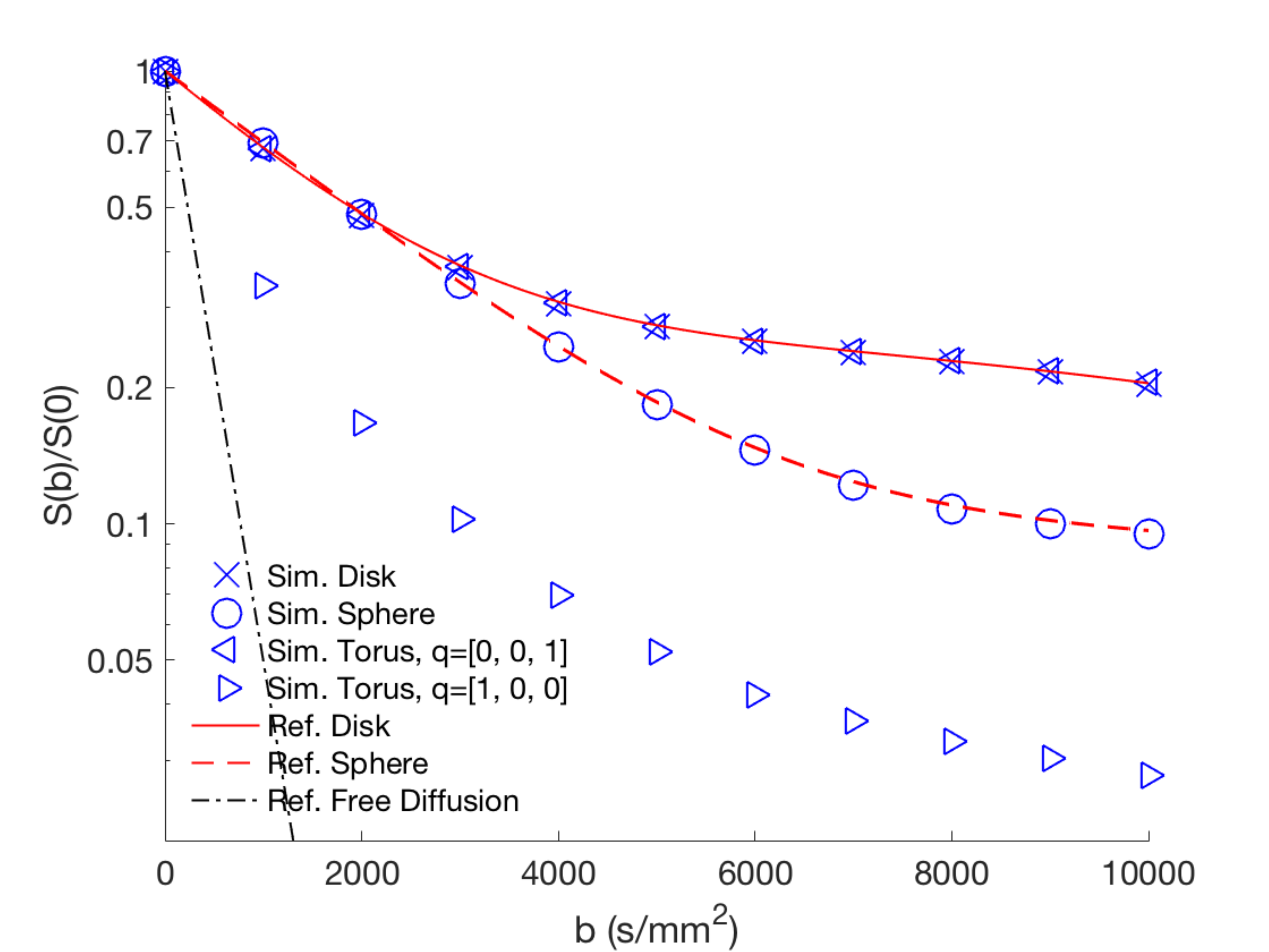}
        \label{fig:multilayered_signals}
    }    
    \caption{Three-layered structures and their signals for a PGSE with $\Delta=43100\tunit, \delta=10600\tunit$. The time step size is $\Delta t=200\tunit$.}
    \label{fig:three_layers}
\end{figure}

We now consider discontinuous diffusion tensors mentioned in Section \ref{sec:diffusion_tensors} to study diffusion in three-layered structures including a disk, a sphere and a torus whose radii are $5, 7.5$ and $ 10\lunit$ respectively (see Figs \ref{fig:circle_3layers}, \ref{fig:sphere_3layers}, and \ref{fig:theta_torus}). For the torus, the radius from the center of the hole to the center of the torus tube is $R=20\lunit$. The simulated signals match very well to the reference signals with the time step size of $\Delta t=200\lunit$ (see Fig.     \ref{fig:multilayered_signals}). The Python source code is available at
\begin{center}
{\small\url{https://colab.research.google.com/github/van-dang/DMRI-FEM-Cloud/blob/master/MultilayeredStructures.ipynb}} 
\end{center}

\begin{figure}[ht]
  \centering
    \subfloat[$T_2=\text{[}\infty, \infty, \infty\text{]}$\label{fig:FEM_T2_inf_inf_inf}]{\includegraphics[width=0.3\textwidth]{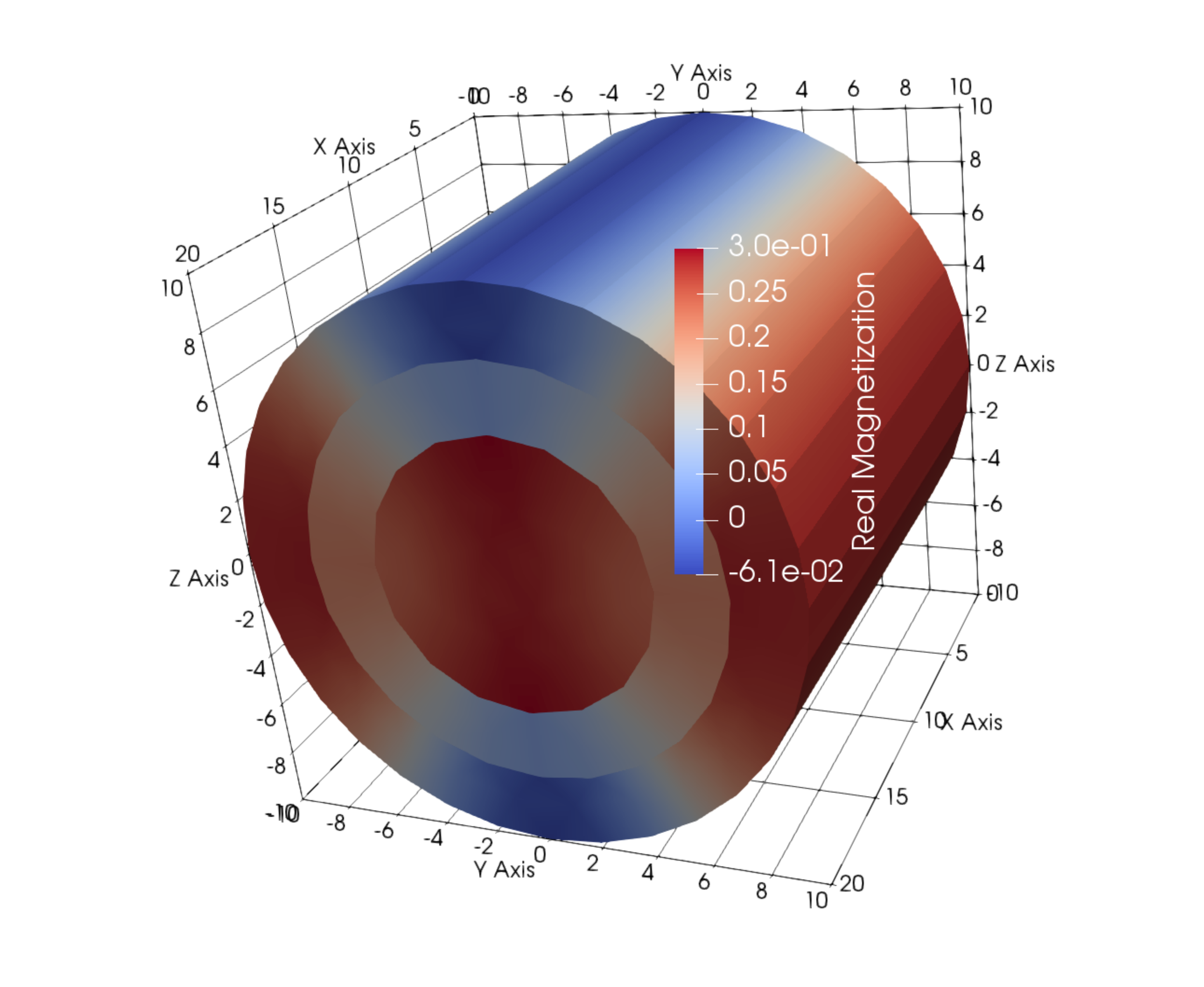}} 
    \subfloat[$T_2=\text{[}40\,{\rm ms}, \infty, \infty\text{]}$\label{fig:FEM_T2_40_inf_inf}]{\includegraphics[width=0.3\textwidth]{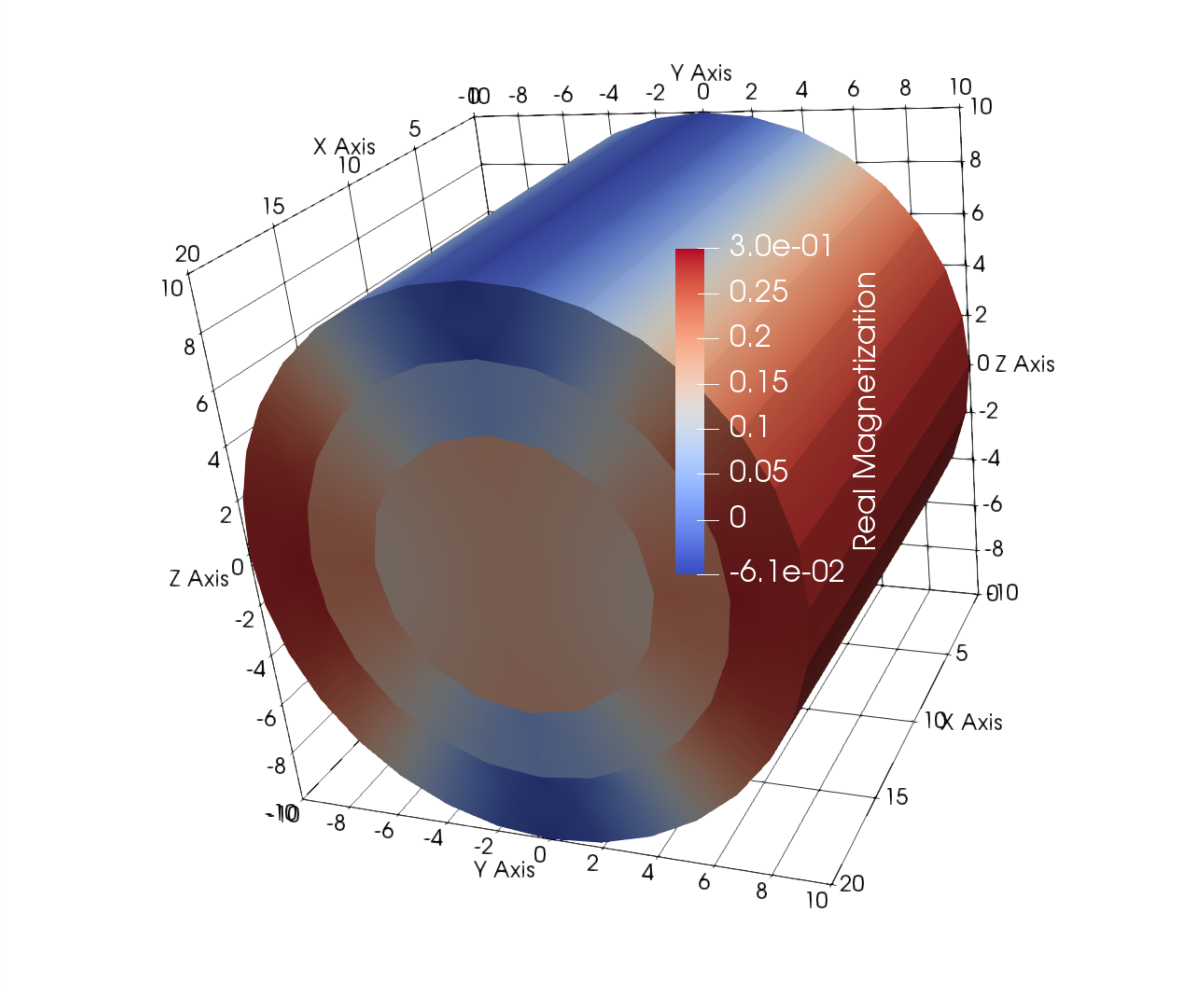}}
    \subfloat[$T_2=\text{[}\infty, 40, 40\text{]\,{\rm ms}}$\label{fig:FEM_T2_inf_40_40}]{\includegraphics[width=0.3\textwidth]{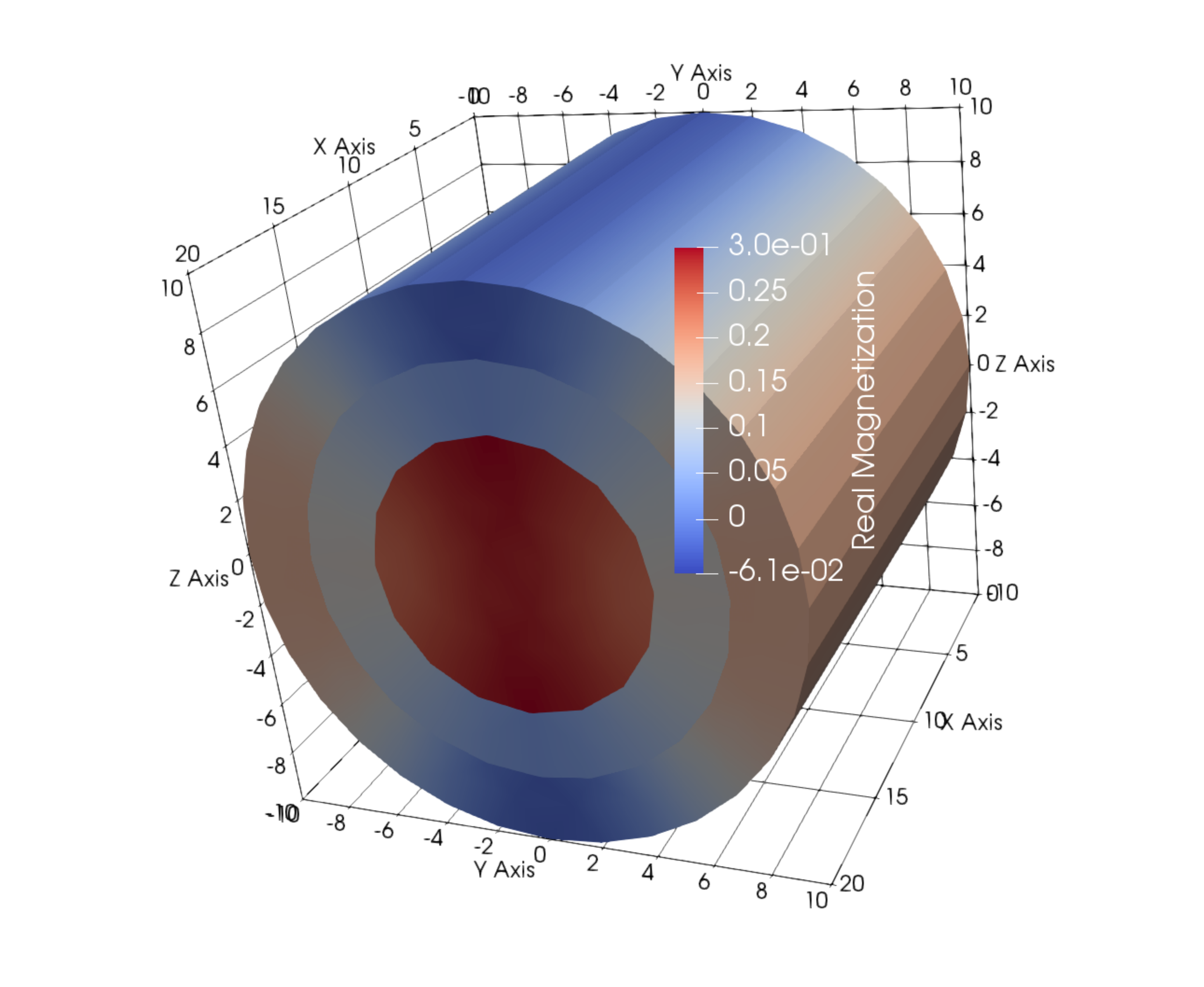}} \\   
    \subfloat[Signals\label{fig:T2_effects}]{\includegraphics[width=1.0\textwidth]{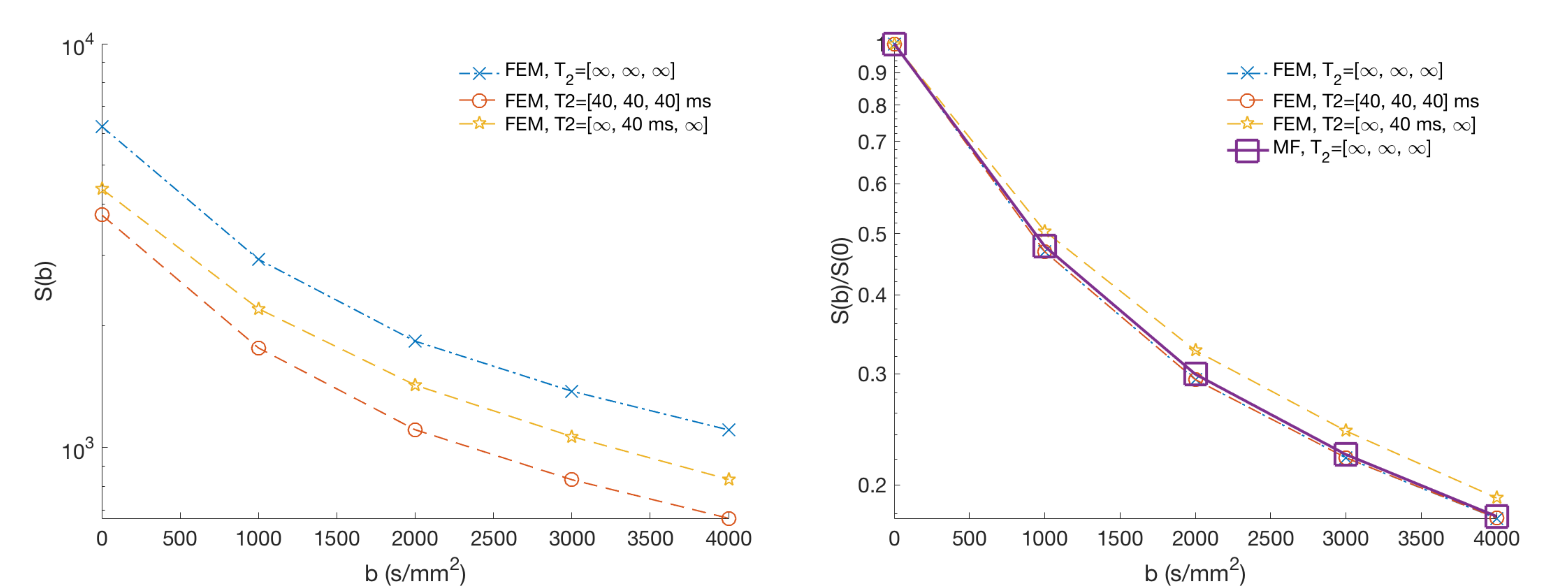}}
  \caption{$T_2-$effects of diffusion inside a three-layered cylinder for a PGSE with $\Delta=\delta=10\rm ms$, permeability $\kappa=10^{-5}\kunit$ to the magnetization at  $b=4000\bunit$ (a, b, c), and to the signals for $b$ between 0 and $4000\bunit$ (d).}
\label{fig:T2_effect_sol}
\end{figure}

The three-layered cylinder is again used to illustrate the effect of $T_2-$relaxation to the magnetization and the signal attenuation. The gradient direction is $\bm q=[0,1,0]$ which is perpendicular to the cylinder axis. As expected, the transverse magnetization decays faster for smaller $T_2$ (Figs. \ref{fig:FEM_T2_inf_inf_inf}, \ref{fig:FEM_T2_40_inf_inf}, \ref{fig:FEM_T2_inf_40_40}). The signals $S(b)$ are quite different when $T_2$ varies (Fig. \ref{fig:T2_effects}). Here we also show that our signals approximate accurately the reference ones calculated by the matrix formalism (solid curve in the figure) \cite{GREBENKOV2010181}. In short, $T_2-$relaxation can be used as one of the sources of the image contrast.
The Python source code is available at 
\begin{center}
{\small\url{https://colab.research.google.com/github/van-dang/DMRI-FEM-Cloud/blob/master/T2_Relaxation.ipynb}}    
\end{center}

\begin{figure}[ht]
  \centering
    \subfloat[Signals \label{fig:signals_diff_profiles_disk_R5}]{\includegraphics[width=0.5\textwidth]{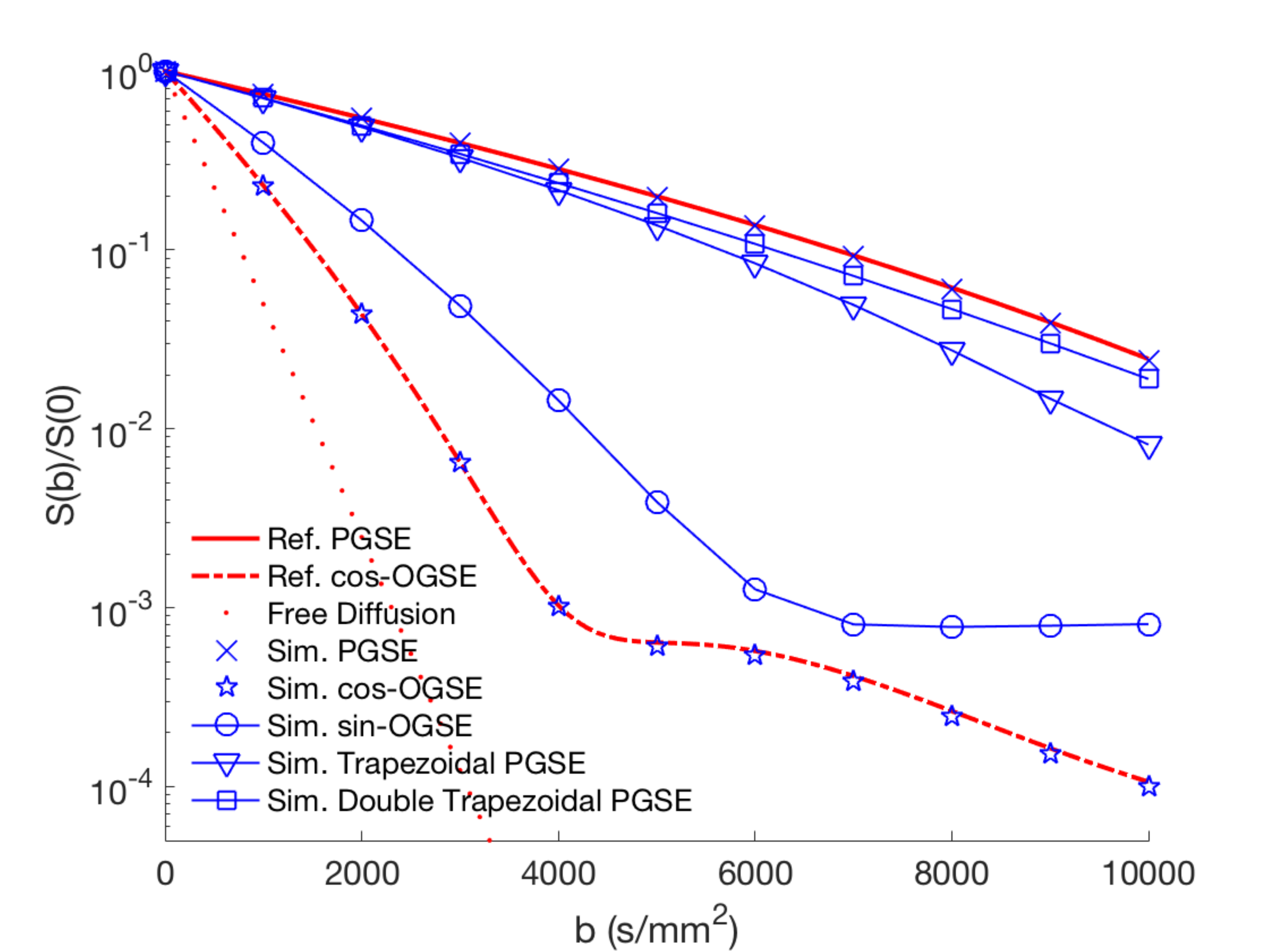}}  
    \subfloat[Temporal profiles \label{fig:arbitrary_temporal_profiles}]{\includegraphics[width=0.5\textwidth]{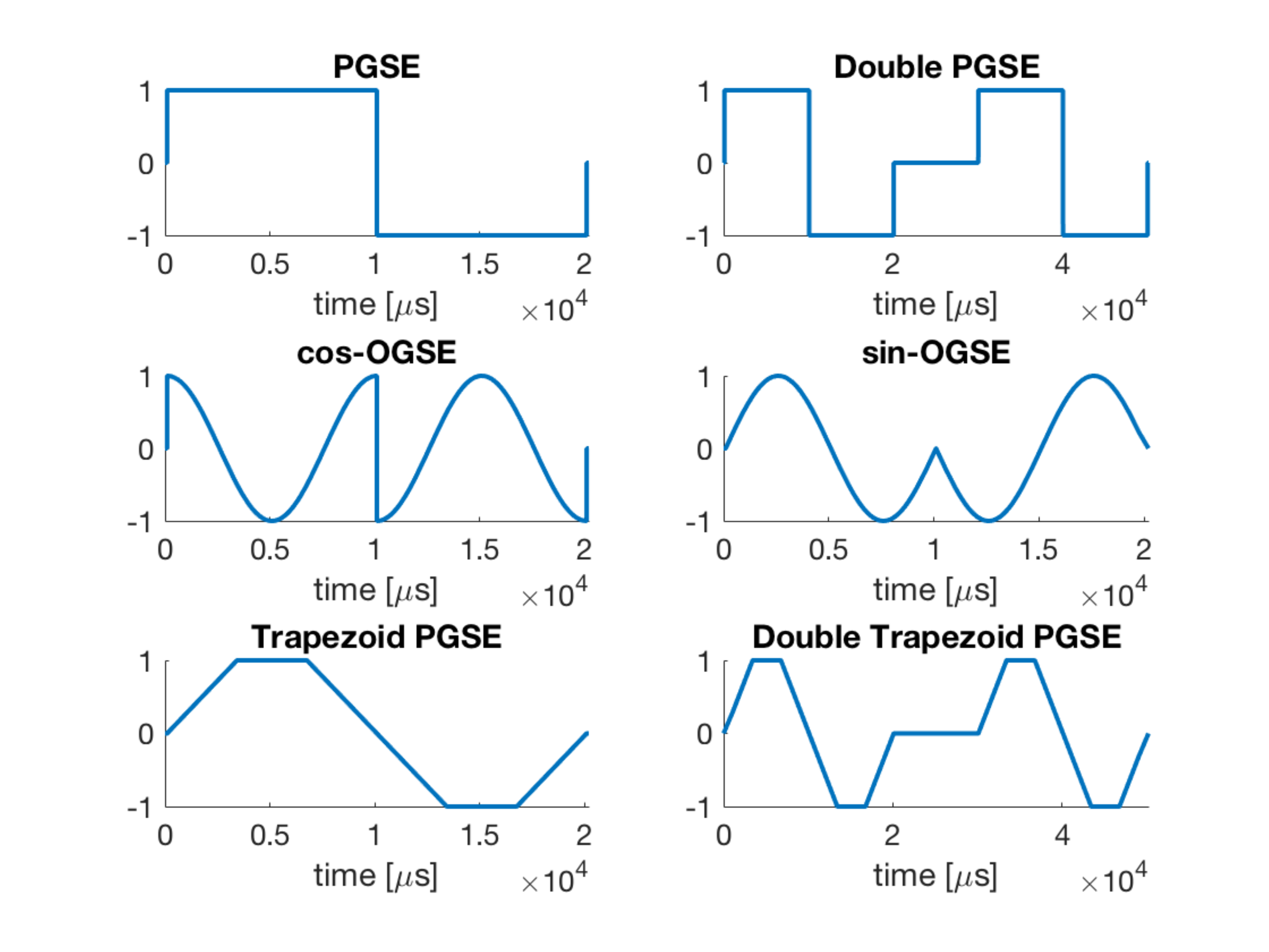}} 
  \caption{Simulated signals inside a disk of radius $5\lunit$ (a) for different temporal profiles: PGSE, Double PGSE, cos-OGSE, sin-OGSE, Trapezoidal PGSE, and Double Trapezoidal PGSE with $\delta=\Delta=10000\tunit$ (b). The simulated signals match very well the reference signals for the PGSE and cos-OGSE. The signals with OGSE sequences decay faster compared to the others.}
\label{fig:arbitrary_profiles}
\end{figure}

Now, we use the solver to compare the signals inside a disk of radius $5\lunit$ for some temporal profiles: PGSE, Double PGSE, cos-OGSE, sin-OGSE, Trapezoidal PGSE, and Double Trapezoidal PGSE with $\delta=\Delta=10000\tunit$ (see Fig. \ref{fig:arbitrary_temporal_profiles}).
The solver is available at
\begin{center}
    {\small\url{https://colab.research.google.com/github/van-dang/DMRI-FEM-Cloud/blob/master/ArbitraryTimeSequence.ipynb}}
\end{center}
The simulated signals match very well the reference signals for the PGSE and cos-OGSE. The signals with OGSE sequences decay faster compared to the others (see Fig. \ref{fig:signals_diff_profiles_disk_R5}).
\section{Simulation examples}\label{sec:examples}
\subsection{Realistic neurons}
We consider a population of 36 pyramidal and 29 spindle neurons. They are distributed in the anterior frontal insula (aFI) and the anterior cingulate cortex (ACC) of the neocortex of the human brain. They share some morphological similarities such as having a single soma and dendrites branching on opposite sides. This population consists of 20 neurons for each type in aFI, and 9 spindles, 16 pyramidals in ACC. We have published these volume meshes at
\begin{center}
{\small\url{https://github.com/van-dang/RealNeuronMeshes}} 
\end{center}

The solver is available at
\begin{center}
    {\small\url{https://colab.research.google.com/github/van-dang/DMRI-FEM-Cloud/blob/master/RealNeurons.ipynb}}
\end{center}

Table \ref{tab:colab_timing} shows the computational time in minutes of neuron simulations on Google Colaboratory for a PGSE sequence with $\Delta=43100\tunit, \delta=10600\tunit$, $b=4000\,\bunit$ and two different time step sizes $\Delta t=50, 100\tunit$. The relative difference in signals between them is within 4\%. With $\Delta=100\tunit$ it costs about an hour for the largest neuron with \num{615146} vertices whereas it costs only 3 minutes with a small neuron with \num{27811} vertices. 
\begin{table}[ht]
\centering
\begin{tabular}{c|c|c|c}
Neuron  & Mesh size & $\Delta t=50\tunit$   & $\Delta t=100\tunit$ \\ \hline\hline
04b\_pyramidal7aACC & \num{615146} vertices & 119 & 64 \\
25o\_spindle17aFI   & \num{51792} vertices & 21 & 10 \\
03b\_pyramidal2aACC & \num{27811} vertices   & 6  & 3
\end{tabular}
\caption{Timing in minutes of neuron simulations on Google Colaboratory for a PGSE sequence with $\Delta=43100\tunit, \delta=10600\tunit$ and different time step sizes $\Delta t$.}
\label{tab:colab_timing}
\end{table}

\begin{figure}[ht]
  \centering
  \subfloat[\label{fig:fru_M_100383}]{
\includegraphics[width=0.45\textwidth]{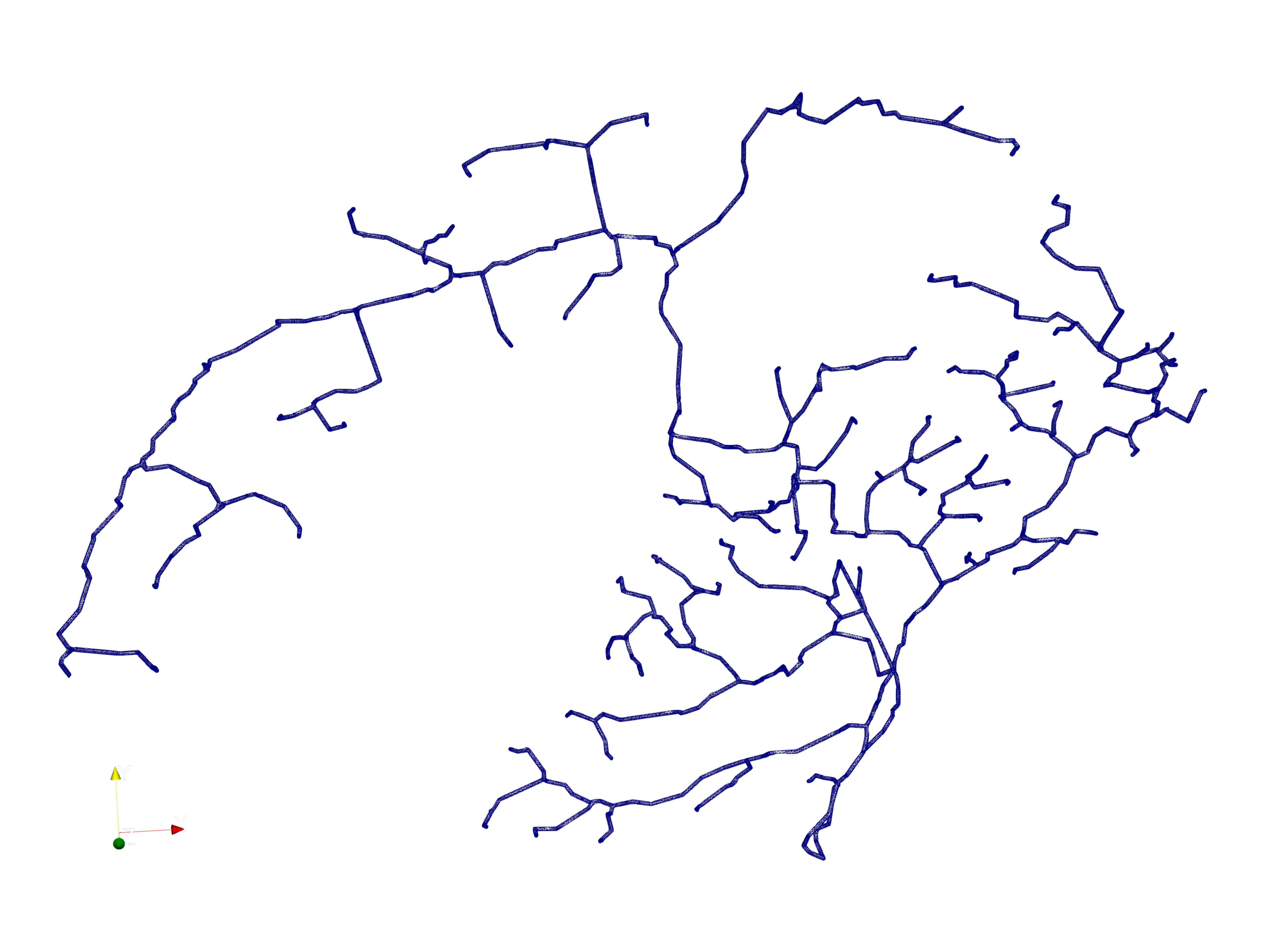}}
  \subfloat[\label{fig:signals_fru_M_100383}]{
\includegraphics[width=0.45\textwidth]{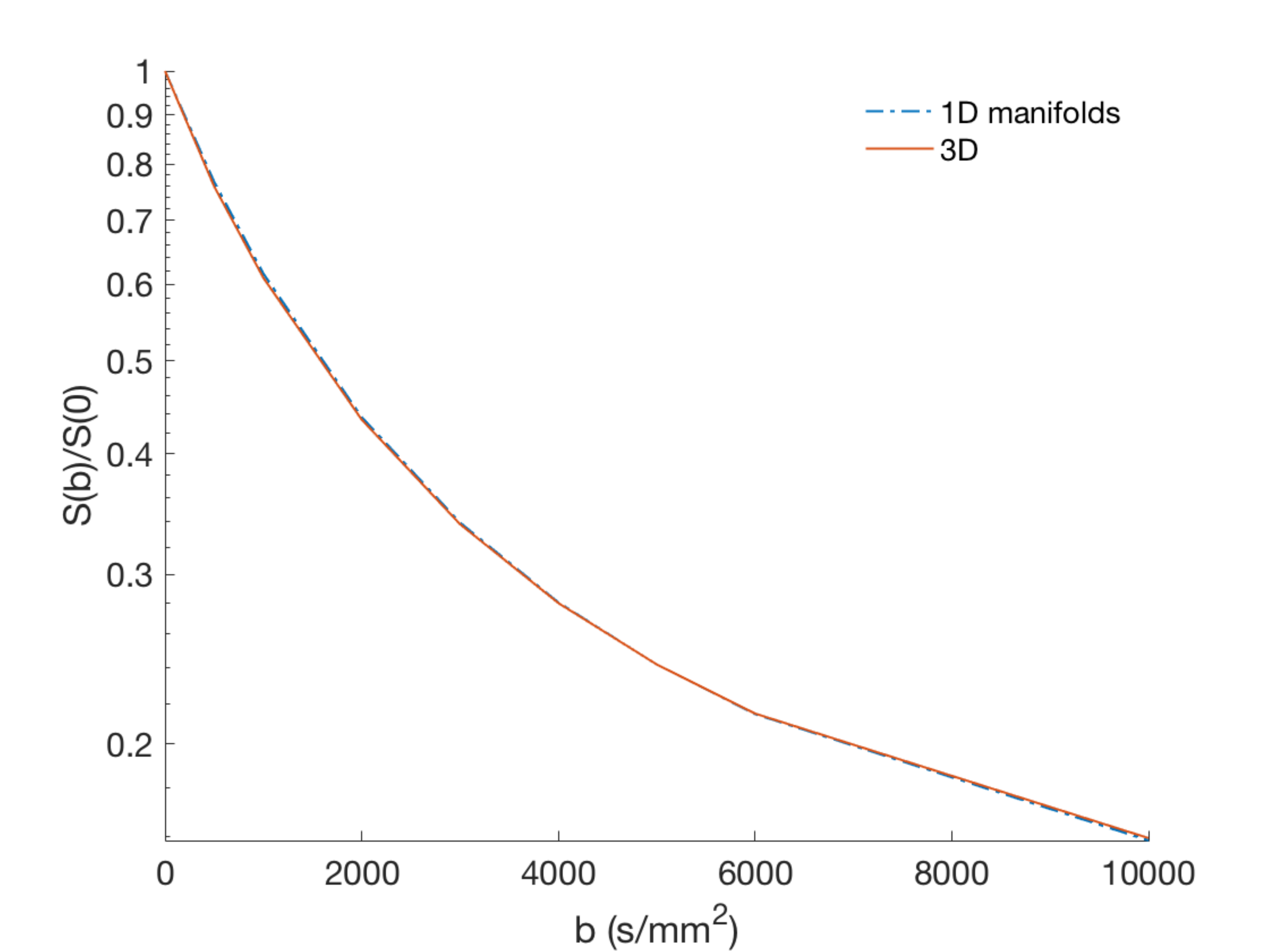}}
  \caption{A comparison between signals inside a neuron from the drosophila melanogaster
for a standard 3D mesh and the corresponding 1D manifolds. For $\Delta t=200\tunit$, it costs only 3 seconds for 1D manifolds but 380 seconds for 3D to compute the signal for one $b-$values with the same accuracy.}
\label{fig:manifolds}
\end{figure}

In addition to the standard approach using volume elements, we also allow for simulating on manifolds following the method developed in \cite{NGUYEN2019176}. 
Figure \ref{fig:manifolds} shows a comparison between signals inside a neuron from the drosophila melanogaster
for a standard 3D mesh and the corresponding 1D manifolds. For $\Delta t=200\tunit$, it costs only 3 seconds for 1D manifolds but 380 seconds for 3D to compute the signal for one $b-$values with the same accuracy.
For more details, it is recommended to look at the solver available at
\begin{center}
    {\small\url{https://colab.research.google.com/github/van-dang/DMRI-FEM-Cloud/blob/master/Manifolds.ipynb}}
\end{center}

\subsection{Extracellular space}
It is challenging to perform simulations on extracellular space (ECS) due to the geometrical complexity. The thickness of ECS is tiny compared to the computational domain. It is extremely time-consuming to use Monte-Carlo approaches. If the reflection condition is applied, the particle undergoes multiple reflections until no further surface intersections are detected, and if the rejection method is applied, the time step sizes need to be very small to be accurate.

\begin{figure}[ht]
  \centering
  \subfloat[\label{fig:2E_ExtraCellular}]{
\includegraphics[width=0.4\textwidth]{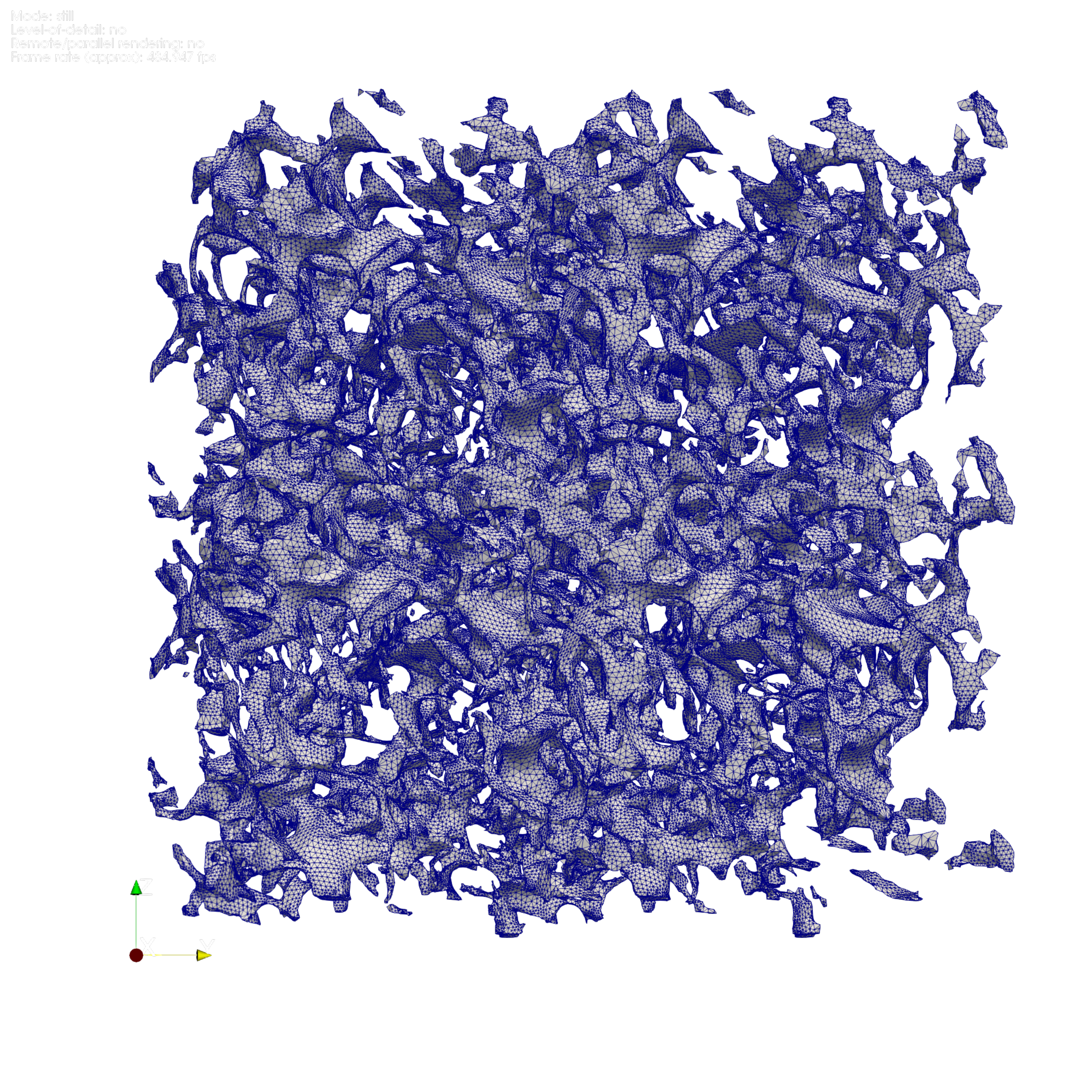}}
  \subfloat[\label{fig:signals_ecs_dt_1000}]{
\includegraphics[width=0.5\textwidth]{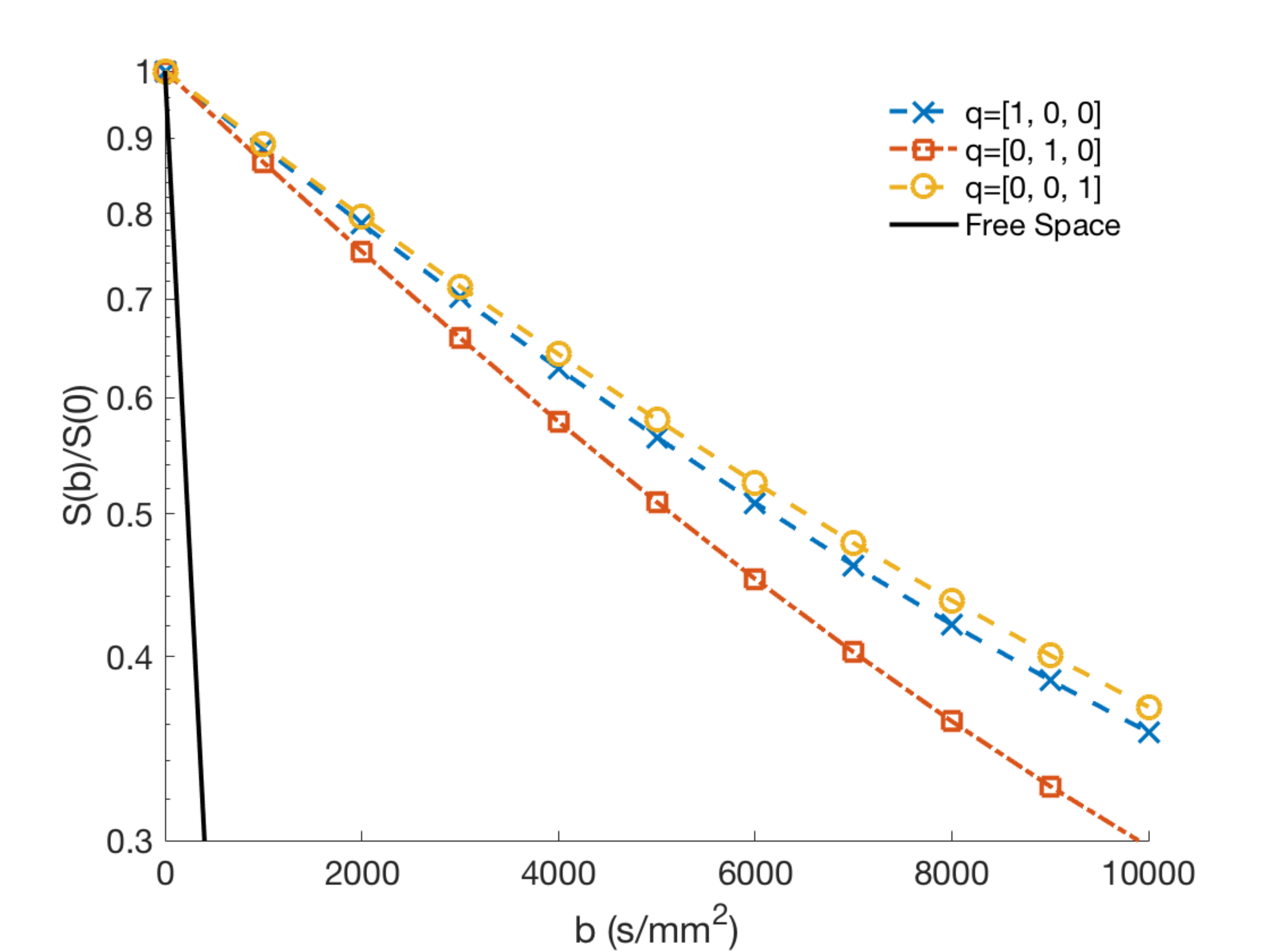}}
  \caption{The mesh of an extracellular space with \num{462420} vertices and \num{926058} tetrahedrons (a). It was reconstructed from the medical segmentation published at \url{http://synapseweb.clm.utexas.edu/2013kinney} (see also \cite{doi:10.1002/cne.23181}). Two directions in the $xz-$plane give quite similar signals showing that the domain is quite isotropic in these directions and they both are distinguishable to the signals in the $y-$direction (b).}
\label{fig:2E_ExtraCellular_group_10um}
\end{figure}

In this section, we show that it is efficient to use our framework. We tested with the ECS extracted from the medical segmentation published at \url{http://synapseweb.clm.utexas.edu/2013kinney} (see also \cite{doi:10.1002/cne.23181}). The volume mesh is shown in Fig. \ref{fig:2E_ExtraCellular} with \num{462420} vertices and \num{926058} tetrahedrons
The processed meshes are available in the following link
\begin{center}
    {\small\url{https://github.com/van-dang/DMRI-FEM-Cloud/raw/mesh/2E_ExtraCellular_group_10um_vol.xml.zip}}
\end{center}

The Google Colab-based solver is available in the following link
\begin{center}
{\small\url{https://colab.research.google.com/github/van-dang/DMRI-FEM-Cloud/blob/master/ExtracellularSpace.ipynb}}
\end{center}
The timing per $b-$value is about 30 minutes on Google Colaboratory for the time discretization $\Delta t=1\,\rm ms$. With half of this time-step size, i.e., $\Delta t=500\,\tunit$, it takes about an hour for $b=10000\bunit$ on Google Colaboratory and the difference in the signals compared to $\Delta t=1000\,\tunit$ is only 1\%. The signals for three principle gradient directions are shown in Fig. \ref{fig:signals_ecs_dt_1000}. Two directions in the $xz-$plane give quite similar signals showing that the domain is quite isotropic in these directions and they both are distinguishable from the signals in the $y-$direction.

\subsection{Parallelization}
Now we verify the simulation performance with the Singularity image on a 12-month free trial of Google Cloud Platform (\url{https://cloud.google.com}) and Tegner (PDC - KTH). The script-based interface is used. It shares the core functionalities with the Python Notebook interface and supports all the functionalities discussed in the paper except the artificial permeability implementation which is still in development. So, the mesh needs to be periodic to have the pseudo-periodic BCs in the parallel execution. 

\begin{figure}[ht]
    \centering
    \subfloat[Neuron \lstinline{04b_pyramidal7aACC}]{
    \includegraphics[width=0.45\textwidth]{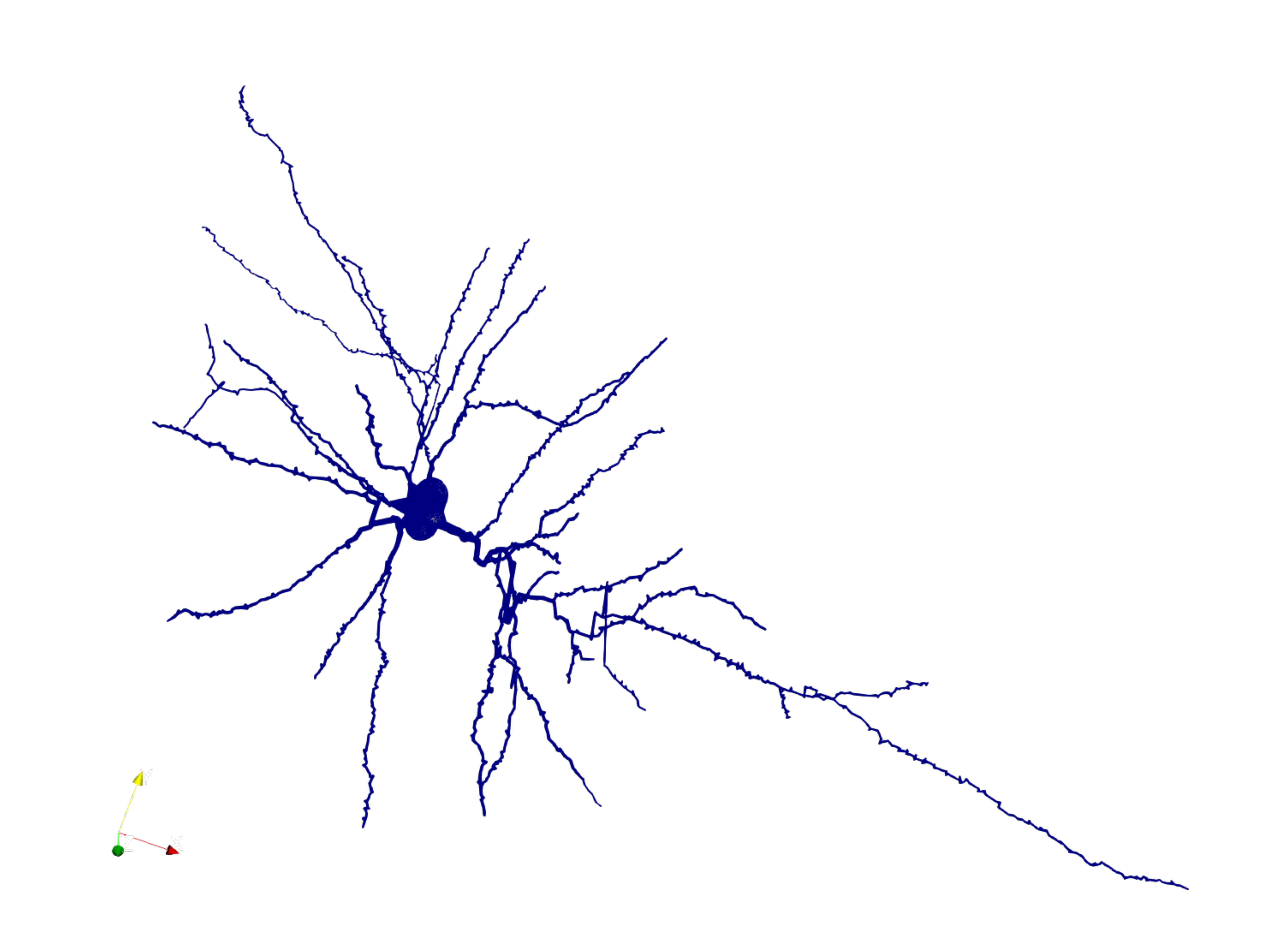}\label{fig:04b_pyramidal7aACC}
    }
    \subfloat[The mouse neuron embedded in a box]{
        \includegraphics[width=0.45\textwidth]{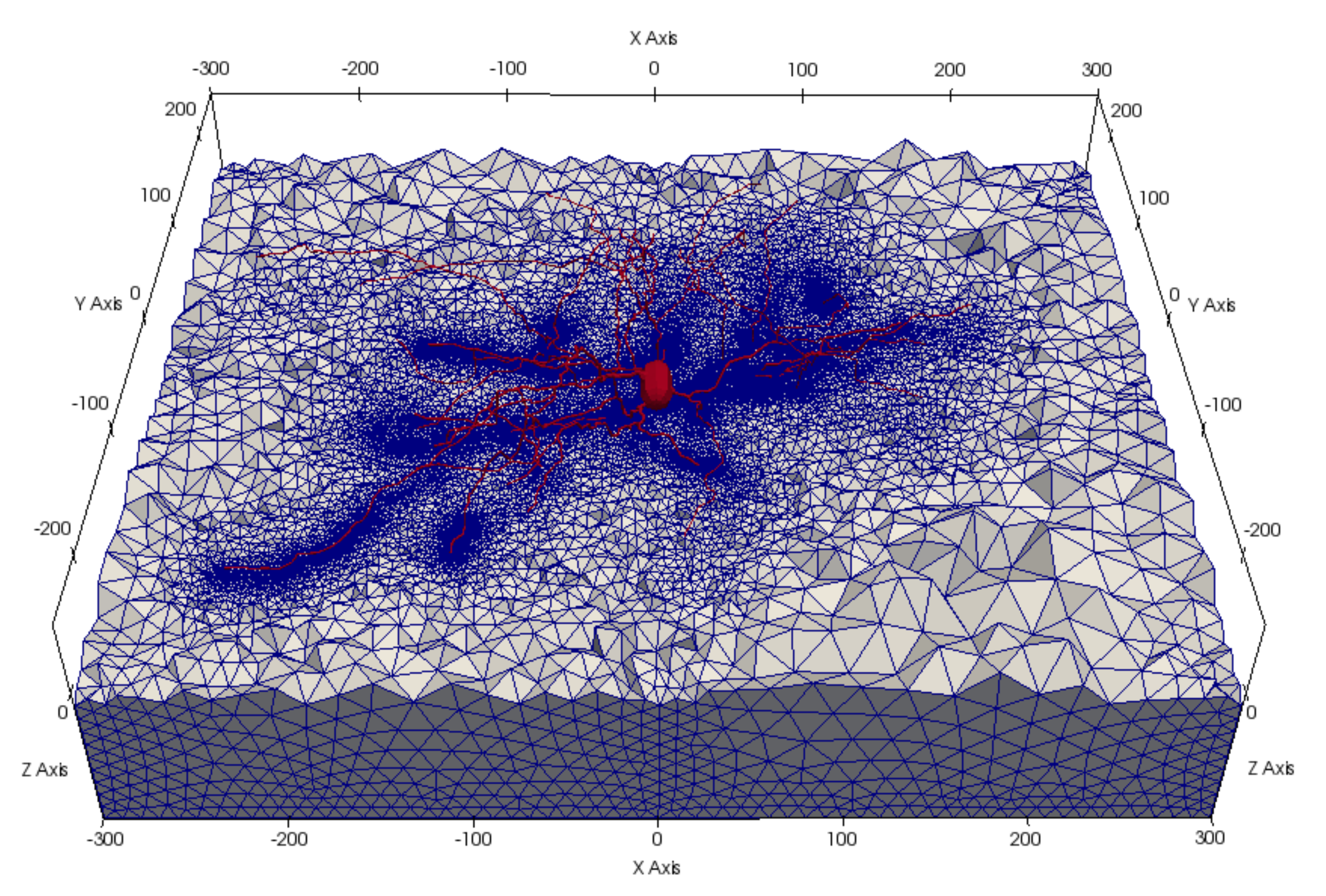}
        \label{fig:computation_domain_neuron}
    }\\
    \subfloat[Strong parallel scaling of one-node and multi-node simulations]{\includegraphics[width=0.7\textwidth]{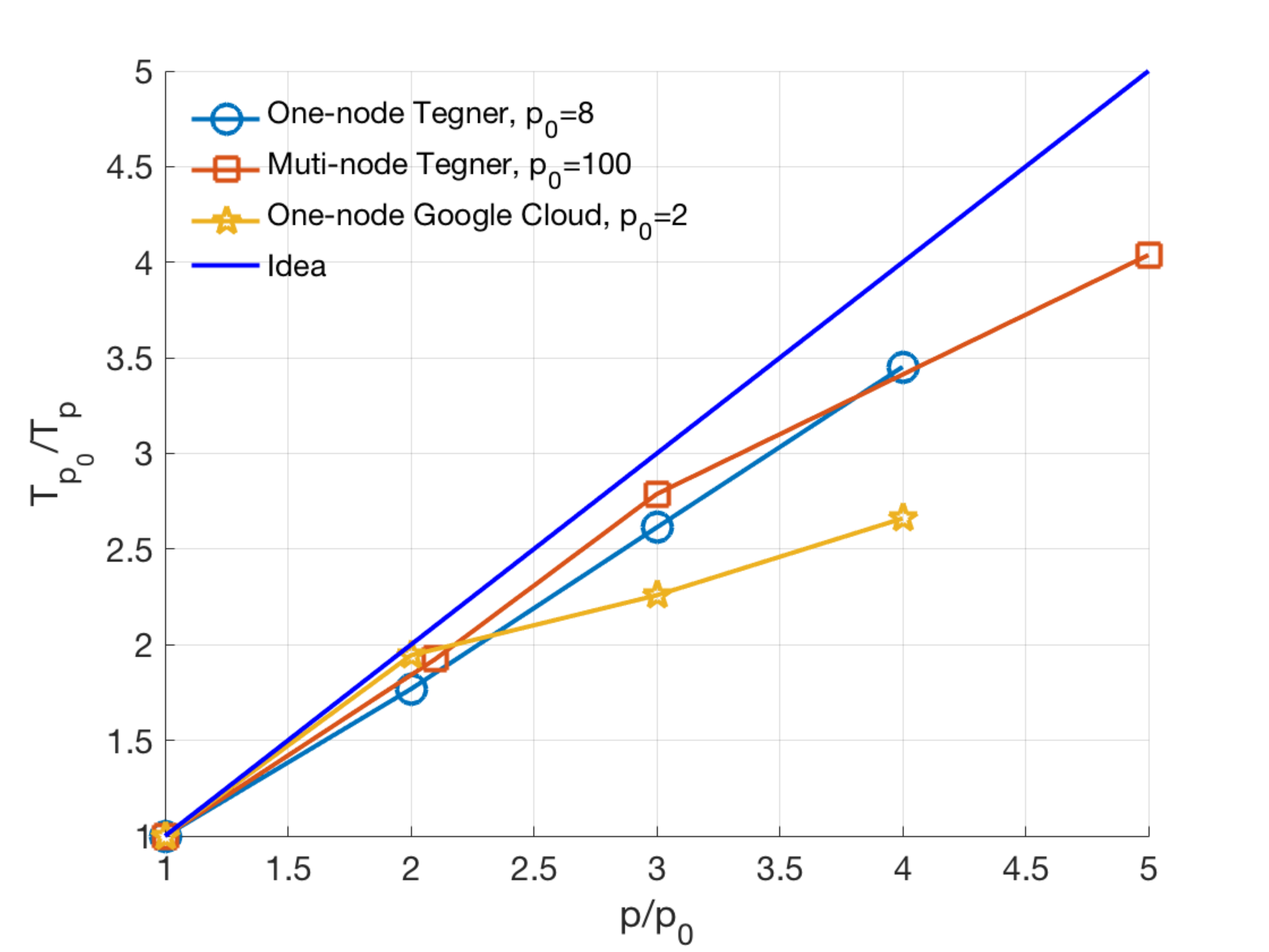}\label{fig:perf_tegner}
    }
    \caption{The simulation performance is verified for a single computational node on the neuron \lstinline{04b_pyramidal7aACC} and for multiples nodes on 
    the mouse neuron embedded in a box presented in Section 7.6 in \cite{NGUYEN2018271}. The strong scaling on Tegner is good  both on one node (32 CPUs) and multi-node (scaling up to 500 CPUs). For a small number of cores (2, 4, 6, 8) on Google Cloud, for this specific case with a fine mesh size of 2.5M tetrahedrons, the work and data partition per process are greatly exceeding the ideal work and data per process ratio. The scaling is less good in comparison with the ideal linear scaling.}
    \label{fig:real_neurons}
\end{figure}

First, we show the simulation performance on one computational node on the neuron \lstinline{04b_pyramidal7aACC} (Fig. \ref{fig:04b_pyramidal7aACC}) with the mesh size of about \num{0.6}M vertices (\num{2.5}M tetrahedrons) and the time discretization of $\Delta t=200\tunit$.  A PGSE with $\Delta=43100\tunit, \delta=10600\tunit$ is used. The commands to execute the simulation with the FEniCS image are follows
\begin{lstlisting}
singularity exec -B $PWD writable_fenics_dmri.simg python3 PreprocessingOneCompt.py -o onecompt_files.h5
ListNumProcs="2  4  6  8"    # one-node in Google Cloud
ListNumProcs="5 10 15 20"   # one-node in Tegner
for p in ${ListNumProcs}
do
    singularity exec -B $PWD writable_fenics_dmri.simg mpirun -n ${p} python3 GCloudDmriSolver.py -f onecompt_files.h5 -M 0 -b 1000 -d 10600 -D 43100 -k 200 -K 3e-3 -gdir 1 0 0
done    
\end{lstlisting}
On Tegner with 20 processors, it costs about 7 minutes per one $b-$value whereas on Google Cloud with 8 processors, it costs about 30 minutes.

Then, we verify with 25 computational nodes on the sample presented in Section 7.6 \cite{NGUYEN2018271}. The sample consists of a pyramidal neuron of an adult female mouse \cite{Carim-Todd678} embedded in the center of a computational domain $\Omega=[-300,300]\times[-250,250]\times[-100,100]\,\lunit^3$ (Fig. \ref{fig:computation_domain_neuron}). We assume that there is a permeable membrane with $\kappa=10^{-5}\kunit$ between the neuron and the extracellular space. The whole mesh (box + neuron) has about \num{1.5}M vertices (\num{8.5}M tetrahedrons). The neuron itself consists of \num{131996} vertices and \num{431326} tetrahedrons. The whole mesh and sub-mesh (neuron) are available for download at
\begin{center}
    {\scriptsize\url{https://github.com/van-dang/DMRI-FEM-Cloud/raw/mesh/volume_box_N_18_7_3_5L_fine.xml.zip}}
    {\scriptsize\url{https://github.com/van-dang/DMRI-FEM-Cloud/raw/mesh/volume_N_18_7_3_5L_fine.xml.zip}}
\end{center}
Below are the commands to execute the simulation with the FEniCS image 
\begin{lstlisting}
singularity exec -B $PWD writable_fenics_dmri.simg python3 PreprocessingMultiCompt.py -o multcompt_files.h5
ListNumProcs="100 210 300 500"   # 25 nodes in Tegner
for p in ${ListNumProcs}
do
    mpirun -n $p singularity exec -B $PWD writable_fenics_dmri.simg python3 GCloudDmriSolver.py -f multcompt_files.h5 -M 1 -b 1000 -p 1e-5 -d 10600 -D 43100 -k 200 -gdir 0 1 0
\end{lstlisting}
The simulation with 500 processors costs about 20 minutes per $b-$value with $\Delta t=200\tunit$.

The strong parallel scaling is shown in Fig.  \ref{fig:perf_tegner}. Here $T_p$ indicates the timing for $p$ processors, $p_0=5$ for the first case and $p_0=100$ for the latter. The scaling for FEniCS on Tegner is good  both on one-node (32 CPUs) and multi-node (scaling up to 500 CPUs). For a small number of cores (2,4,6,8) on Google Cloud, for this specific case with a fine mesh size of 2.5M tetrahedrons, the work and data partition per process are greatly exceeding the ideal work and data per process ratio. The scaling is less good in comparison with the ideal linear scaling.

\section{Discussion}\label{sec:discussion_future_work}
The proposed framework can be viewed as the Python version of the Matlab-based SpinDoctor.  In this paper, we focused on advanced software features such as portability and parallelization.  This framework inherits all of the PDE solution functionalities of FEniCS, thus, extensions and generalizations of the present dMRI simulation problem, including the coupling with flow, the simulation on deforming domains like the heart, or the coupling of simulations in manifolds with simulations in 3D domains, are rather straightforward.

Similar to SpinDoctor, the present framework is supposed to be faster and more accurate than Monte-Carlo simulation packages such as Camino. More importantly, our approach benefits from a long history of theoretical and numerical developments by the mathematical and engineering communities.   It enhances software reliability which is one of the core concerns in medical applications.

As other cloud-based software developments, this framework brings reproducible science and open-source software to computational diffusion MRI.  It speeds up the method development process since the results are easy to confirm and new methods can be easily developed on top of the existing methods. New algorithms written as Google Colaboratory notebooks can quickly circulate in the MRI community and this allows for active collaboration between research groups.

Since SpinDoctor couples the finite elements discretization with optimized adaptive ODE solvers, it is more efficient than our framework in terms of time discretization. The analogous ODE solvers written in Python can be found in the SciPy Library \cite{scipy} but they are not ready to use within our framework: they do not yet efficiently support the mass matrix and the sparse Jacobian matrix. The lumped mass matrix approach can be used to fix the first issue but more investigations are needed to resolve the latter issue.

Generating finite elements meshes from medical segmentation is very challenging.  Complicated surface meshes currently need to be processed outside the framework to obtain a good quality finite elements mesh.  Streamlining this process is an interesting direction of future investigation and it may be well worthwhile to develop algorithms to automate this process.
\section{Conclusions}\label{sec:conclusion}
We proposed a portable simulation framework for computational diffusion MRI that works efficiently with cloud technology. The framework can be seamlessly integrated with cloud computing resources such as Google Colaboratory notebooks working on a web browser or with Google Cloud Platform with MPI parallelization. Many simulation needs of the field were addressed by the use of advanced finite element methods for both single- and muti-compartment diffusion domains, with or without permeable membrane and periodic boundaries.  We showed the accuracy, the computational times, and parallel computing capabilities through a set of examples, while mentioning straightforward future extensions. The framework contributes to reproducible science and open-source software in computational diffusion MRI.  We hope that it will help to speed up method developments and stimulate research collaborations.
\section*{Acknowledgement}
This research has been supported by the Swedish Energy Agency, Sweden with the project ID P40435-1; MSO4SC with the Grant No. 731063; the Basque Excellence Research Center (BERC 2014–2017) program by the Basque Government; the Spanish Ministry of Economy and Competitiveness MINECO: BCAM Severo Ochoa accreditation SEV-2013-0323; the ICERMAR ELKARTEK project of the Basque Government; the projects of the Spanish Ministry of Economy and Competitiveness with reference MTM2013-40824-P and MTM2016-76016-R. The research was conducted on resources provided by the Swedish National Infrastructure for Computing (SNIC) at the Center of High- Performance Computing (PDC). We also would like to thank ANSA from Beta-CAE Systems S. A., who generously provided an academic license. The first author would like to thank Niyazi Cem Degirmenci for his enthusiastic supports.
\appendix
\section*{Appendices}
 The methods imposed in our framework are based on the partition of unity finite element method (PUFEM) to manage the interface conditions \cite{NGUYEN2018271}. Both weak and strong implementation of the periodicity with some advantages and disadvantages are included. The $\theta-$method is used for the time discretization.

\section{Strong implementation of the pseudo-periodic BCs}\label{sec:strong_bc_fem}

 The complex-valued and time-dependent term in the pseudo-periodic boundaries make it too difficult to implement in a standard FEM software package. So, one can transform the pseudo-periodic BCs to the periodic ones. Following \cite{Russell2012, Nguyen2014283}, one can choose to transform the magnetization
to a new unknown $u(\bx,t)$:
\begin{equation*}
u(\bx,t)=U(\bx,t)\,e^{i\,\,\gamma\,\mathcal{F}(t)\,\bm g\cdot\bx}
\end{equation*}
The Bloch-Torrey PDE (\ref{eq:strong_BT}) is then transformed to \cite{Nguyen2014283}
\begin{equation}
	\frac{\partial u}{\partial t} =-i\,\gamma \,\mathcal{F}\Bigl(\bm g\cdot {\bm D}\nabla u+\nabla u\cdot {\bm D}\,\bm g\Bigl)-\Bigl(\gamma\,\mathcal{F}\Bigl)^2 {\bm g}\cdot \bm{D}\,{\bm g}\,u -\frac{u}{T_2}
+\nabla \cdot \Bigl(\bm{D}\nabla u\Bigl),
	\label{eqn:trBlo}
\end{equation}
with periodic BCs
\begin{equation}\label{Periodic}
	\begin{aligned}
		u_m&=u_s\\
		{\bm D}_m\nabla u_m\cdot \bm n&={\bm D}_s\nabla u_s\cdot \bm n
	\end{aligned}
\end{equation}
The homogeneous Neumann boundary condition of $U$ leads to
\begin{equation*}
	\bm{D}\nabla u\cdot\bm{n} = i\,\gamma\,\mathcal{F}\,u\, 
    \bm{D}\,\bm g\cdot\bm{n}.
\end{equation*}
The interface conditions (Eq. \ref{eq:btpde_ic}) are
changed to
\begin{equation}\label{Binterface2}
\begin{aligned}
    \Bigl\llbracket \bm{D} \nabla u \cdot \bm n_0 \Bigl\rrbracket &=  2\,i\,\gamma\,\mathcal{F}\Bigl\{u\, 
    \bm{D}\,\bm g\cdot\bm{n}_0\Bigl\}.\\
    \Bigl\{\bm{D} \nabla u \cdot \bm n_0 \Bigl\} &= -\kappa \llbracket u\rrbracket+\frac{i\,\gamma\,\mathcal{F}}{2}\llbracket u\, \bm{D}\,\bm g\cdot\bm{n}_0 \rrbracket.
\end{aligned}
\end{equation}
Since the magnetization is discontinuous ($m_0\neq m_1$ on the interface), Eq. (\ref{Binterface2}) shows the flux is also discontinuous.

Following the same PUFEM approach proposed in \cite{NGUYEN2018271}, we obtain the following weak form 
\begin{equation*}
	\Biggl(\frac{\partial}{\partial t} u,v\Biggl)_{\Omega_0\cup \Omega_1} =F(u,v,t).
	\label{eqn:weaktrBlo}
\end{equation*}
where
\begin{multline}
    F(u,v,t)=-\Biggl( i\,\gamma \,\mathcal{F}\,\Bigl({\bm g}\cdot \bm{D}\,\nabla u+\nabla u\cdot \bm{D}\,{\bm g}\Bigl),v\Biggl)_{\Omega_0\cup \Omega_1}\\
	-\Biggl(\bigl(\gamma\,\mathcal{F}\bigl)^2\,{\bm g}\cdot \bm{D}\,{\bm g}\, u + \frac{u}{T_2}, v\Biggl)_{\Omega_0\cup \Omega_1}
	-\Biggl( \bm{D}\nabla u, \nabla v \Biggl)_{\Omega_0\cup \Omega_1}\\
	+\Biggl<-\kappa \llbracket u \rrbracket+\frac{i\,\gamma\,\mathcal{F}}{2}\Bigl\llbracket u\, 
    \bm{D}\,\bm g\cdot\bm{n}_0\Bigl\rrbracket,  \llbracket v \rrbracket\Biggl>_{\Gamma} 
    + \Biggl<2\,i\,\gamma\,\mathcal{F}\Bigl\{ u\, 
    \bm{D}\,\bm g\cdot\bm{n}_0\Bigl\},  \{ v \}\Biggl>_{\Gamma} \\
    +\Biggl< i\,\gamma\,\mathcal{F}\,u\, 
    \bm{D}\,\bm g\cdot\bm{n}, v \Biggl>_{\Gamma^N_0\cup \Gamma^N_1}.
\end{multline}

We consider a partition of the time domain $0=t_0<t_1<\dots<t_N=T$ associated with the time intervals $I_n=(t^{n-1}, t^n]$ of length $k^n=t^n-t^{n-1}$ and $u^n$ be an approximation of $u(\bm x, t)$ for a given a triangulation $\mathcal{T}^h$ at $t=t^n$. 

The PUFEM with the time-stepping $\theta-$method is stated as: Find $u^n_h=(u^n_{h,0}, u^n_{h,1})\in \bm V_h$ such that
\begin{equation}\label{eq:theta_discretization}
\Biggl(\frac{u^n_h-u^{n-1}_h}{k^n}, v_h\Biggl)_{\Omega_0\cup\Omega_1}=\theta\, F(u^n_h, v_h, t^n) + (1-\theta)\, F(u^{n-1}_h, v_h, t^{n-1})
\end{equation}
for all $v_h=(v_{0,h}, v_{1,h})\in \bm V_h$, where
$$\Bigl(a, b\Bigl)_{\Omega_{0,h}\cup\Omega_{1,h}} = \Bigl((1-\Phi_h)\,a_0, b_0\Bigl)_{\Omega_h}+\Bigl(\Phi_h a_1, b_1\Bigl)_{\Omega_h},$$ and $\Phi_h$ is an element-wise constant function:
\begin{equation}\label{eq:phase_func}
    \Phi_h = \begin{cases}
    1 & \mbox{ in }\Omega_{1,h}\\
    0 & \mbox{ in }\Omega_{0,h}
    \end{cases}
\end{equation}
The bilinear and linear forms are defined by
\begin{equation}\label{eq:linear_billinear}
\begin{aligned}
    \bm a(u^n_h,v_h)&=\Biggl(\frac{u^n_h}{k^n}, v_h\Biggl)_{\Omega_0\cup\Omega_1}-\theta\,F(u^n_h,v_h,t^n)\\
    \bm L(v_h)&=\Biggl(\frac{u^{n-1}_h}{k^n}, v_h\Biggl)_{\Omega_{0,h}\cup\Omega_{1,h}}+(1-\theta)\,F\biggl(u^{n-1}_h,v_h,t^{n-1}\biggl)
\end{aligned}
\end{equation}

\section{Weak implementation of the pseudo-periodic BCs}

The pseudo-periodic BCs (Eq. \ref{eq:btpde_bc}) can be 
implemented weakly through the use of an artificial permeability coefficient,
 $\kappa^e$ \cite{Nguyen1080573, NGUYEN2018271}.  The artificial permeability condition at the external boundaries take two equations for the master side and the slave side of the mesh. For the master side, it has the following form
\begin{equation}\label{eq:bjump1}
      \bm{D}_m\nabla U_m \cdot \bm n_m= \kappa^{e} \Bigl(U_{s}\,e^{i\,\theta_{ms}}- U_m\Bigl),
\end{equation}
and for the slave-side it has the following form
\begin{equation}\label{eq:bjump2}
      \bm{D}_s\nabla u_s \cdot \bm n_s= \kappa^{e} \Bigl(U_{m}\,e^{i\,\theta_{sm}}- U_s\Bigl),
\end{equation}
where $U_s=U(\bm x_s), U_m=U(\bm x_m)$, $\theta_{ms}=-\theta_{sm}=\gamma \;\bm g\cdot(\bm x_s-\bm x_m) \,\mathcal{F}(t)$. When the master side is considered (Eq. \ref{eq:bjump1}), $\bm x_m$ is the mesh point and $\bm x_s$ is the projection of $\bm x_m$ onto the slave side. Similarly, when the slave side is considered (Eq. \ref{eq:bjump2}), $\bm x_s$ is the mesh point and $\bm x_m$ is the projection of $\bm x_s$ onto the master side. So, the points always align each other but they do not need to be the mesh grid at the same time. So, this method allows for non-matching meshes.

The artificial permeability coefficient $\kappa^e$ can be chosen to be consistent with the Nitsche's method for the Dirichlet BCs \cite{Nitsche1971} (see also a review in \cite{GAMM:GAMM201490018} and references therein), i.e
$\kappa^e=\max\Bigl\{\frac{\bm D}{h}\Bigl\}$ where $h$ is the element size.

To overcome the CFL constraints, the following operator splitting can be used to have an unconditionally stable scheme 
\begin{equation}\label{eq:weak_periodic}
	\begin{aligned}
 	     \bm{D}_m\nabla U_m \cdot \bm n_m\approx  \kappa^{e} \Bigl(U_s^{n-1}\,e^{i\,\theta_{ms}^n}- U_m^n\Bigl),\\
 	     \bm{D}_s\nabla U_s \cdot \bm n_s \approx \kappa^{e} \Bigl(U_m^{n-1}\,e^{i\,\theta_{sm}^n}- U_s^n\Bigl).
	\end{aligned}
\end{equation}
where $U^n$ and $U^{n-1}$ are the approximations at the current and previous time step respectively.

Without imposing the weak pseudo-periodic, the PUFEM with the time-stepping $\theta-$method is stated as: Find $U^n=(U^n_0, U^n_1)\in \bm V^h$ such that
\begin{equation}\label{eq:theta_discretization2}
\Biggl(\frac{U^n-U^{n-1}}{k^n}, v^h\Biggl)_{\Omega_0\cup\Omega_1}=\theta\, F(U^n, v^h, t^n) + (1-\theta)\, F(U^{n-1}, v^h, t^{n-1})
\end{equation}
for all $v^h=(v^h_0, v^h_1)\in \bm V^h$,
where 
\begin{equation}\label{eq:loadvec}
    F(U,v,t)=\Biggl(-i\,\gamma f(t)\, \bm g \cdot \bm x \, U-\frac{U}{T_2}, v\Biggl)_{\Omega_0\cup\Omega_1}- \Bigl( \bm D \, \nabla U,\nabla v \Bigl)_{\Omega_0\cup\Omega_1}-\kappa\Bigl<\llbracket U \bigl\rrbracket,\llbracket v \bigl\rrbracket  \Bigl>,
\end{equation}
and
$\Bigl(a, b\Bigl)_{\Omega_0^h\cup\Omega_1^h} = \Bigl((1-\Phi^h)\,a_0, b_0\Bigl)_{\Omega^h}+\Bigl(\Phi^h a_1, b_1\Bigl)_{\Omega^h}$, $\Phi^h$ is an element-wise constant function.

The bilinear and linear forms are defined by
\begin{equation}\label{eq:linear_billinear2}
\begin{aligned}
    \bm a(U^n,v^h)&=\Biggl(\frac{U^n}{k^n}, v^h\Biggl)_{\Omega_0\cup\Omega_1}-\theta\,F(U^n,v^h,t^n)\\
    \bm L(v^h)&=\Biggl(\frac{U^{n-1}}{k^n}, v^h\Biggl)_{\Omega_0\cup\Omega_1}+(1-\theta)\,F\biggl(U^{n-1},v^h,t^{n-1}\biggl)
\end{aligned}
\end{equation}

The linear system of equations corresponding to the bilinear and linear forms (Eq. \ref{eq:linear_billinear2}) is
\begin{equation}\label{eq:matrix_form}
    \bm A \,{\bm U^n} = \bm F
\end{equation}
where 
\begin{equation}
   \bm A=\bm M\,{(k^n)^{-1}}-\theta\Biggl(-\Bigl(i\,\gamma\,f^n +\frac{1}{T_2}\Bigl)\, \bm J - \bm S - \bm I \Biggl)
\end{equation}
Here $\bm M$ and $\bm S$ are referred to as the mass and stiffness matrices respectively, $\bm J$ and $\bm I$ are corresponding to the first and third terms on the right-hand side of $F$ (Eq. \ref{eq:loadvec}), i.e  $(\bm g \cdot \bm x \, U, v)$ and $\kappa\Bigl<\llbracket U \bigl\rrbracket,\llbracket v \bigl\rrbracket  \Bigl>$. 

To impose the weak periodic BCs, we plug Eq. (\ref{eq:weak_periodic}) to the linear and bilinear forms

\begin{equation*}
    \bm a^*(U^n_h,v_h) = a(U^n_h,v_h) +\theta\kappa^e\Biggl( \Bigl< U^n_h,v_h\Bigl>_{\Gamma_m^0\cup\Gamma_m^1} + \Bigl< U^n_h,v_h\Bigl>_{\Gamma_s^0\cup\Gamma_s^1}\Biggl)
\end{equation*}

\begin{equation*}
    \bm L^*(v_h) = L(v_h) +(1-\theta)\kappa^e\Biggl( \Bigl< U^{n-1}_{s,h}\,e^{i\,\theta_{ms}^n},v_h\Bigl>_{\Gamma_m^0\cup\Gamma_m^1} + \Bigl< U^{n-1}_{m, h}\,e^{i\,\theta_{sm}^n},v_h\Bigl>_{\Gamma_s^0\cup\Gamma_s^1}\Biggl)
\end{equation*}


\bibliography{ref.bib}

\begin{thebibliography}{10}
\expandafter\ifx\csname url\endcsname\relax
  \def\url#1{\texttt{#1}}\fi
\expandafter\ifx\csname urlprefix\endcsname\relax\def\urlprefix{URL }\fi
\expandafter\ifx\csname href\endcsname\relax
  \def\href#1#2{#2} \def\path#1{#1}\fi

\bibitem{nla.cat-vn2111911}
B.~D. Hughes, Random walks and random environments / Barry D. Hughes, Clarendon
  Press Oxford ; New York, 1995.

\bibitem{Yeh2013}
C.-H. Yeh, B.~Schmitt, D.~Le~Bihan, J.-R. Li-Schlittgen, C.-P. Lin, C.~Poupon,
  \href{http://www.ncbi.nlm.nih.gov/pmc/articles/PMC3794953/}{Diffusion
  microscopist simulator: A general monte carlo simulation system for diffusion
  magnetic resonance imaging}, PLoS One 8~(10) (2013) e76626,
  pONE-D-13-18755[PII].
\newblock \href {http://dx.doi.org/10.1371/journal.pone.0076626}
  {\path{doi:10.1371/journal.pone.0076626}}.
\newline\urlprefix\url{http://www.ncbi.nlm.nih.gov/pmc/articles/PMC3794953/}

\bibitem{4797853}
M.~G. Hall, D.~C. Alexander, Convergence and parameter choice for monte-carlo
  simulations of diffusion mri, IEEE Transactions on Medical Imaging 28~(9)
  (2009) 1354--1364.
\newblock \href {http://dx.doi.org/10.1109/TMI.2009.2015756}
  {\path{doi:10.1109/TMI.2009.2015756}}.

\bibitem{Palombo2016}
M.~Palombo, C.~Ligneul, C.~Najac, J.~Le~Douce, J.~Flament, C.~Escartin,
  P.~Hantraye, E.~Brouillet, G.~Bonvento, J.~Valette,
  \href{http://www.ncbi.nlm.nih.gov/pmc/articles/PMC4914152/}{New paradigm to
  assess brain cell morphology by diffusion-weighted mr spectroscopy in vivo},
  Proc Natl Acad Sci U S A 113~(24) (2016) 6671--6676, 201504327[PII].
\newblock \href {http://dx.doi.org/10.1073/pnas.1504327113}
  {\path{doi:10.1073/pnas.1504327113}}.
\newline\urlprefix\url{http://www.ncbi.nlm.nih.gov/pmc/articles/PMC4914152/}

\bibitem{VANNGUYEN2018}
K.~V. Nguyen, E.~H. Garzon, J.~Valette,
  \href{http://www.sciencedirect.com/science/article/pii/S1090780718302386}{Efficient
  gpu-based monte-carlo simulation of diffusion in real astrocytes
  reconstructed from confocal microscopy}, Journal of Magnetic Resonance\href
  {http://dx.doi.org/https://doi.org/10.1016/j.jmr.2018.09.013}
  {\path{doi:https://doi.org/10.1016/j.jmr.2018.09.013}}.
\newline\urlprefix\url{http://www.sciencedirect.com/science/article/pii/S1090780718302386}

\bibitem{Cook2006CaminoOD}
P.~A. Cook, Y.~Bai, S.~Nedjati-Gilani, K.~K. Seunarine, M.~G. Hall, G.~J.~M.
  Parker, D.~C. Alexander, Camino: Open-source diffusion-mri reconstruction and
  processing, 2006.

\bibitem{Hwang2003}
S.~N. Hwang, C.-L. Chin, F.~W. Wehrli, D.~B. Hackney,
  \href{http://dx.doi.org/10.1002/mrm.10536}{An image-based finite difference
  model for simulating restricted diffusion}, Magnetic Resonance in Medicine
  50~(2) (2003) 373--382.
\newblock \href {http://dx.doi.org/10.1002/mrm.10536}
  {\path{doi:10.1002/mrm.10536}}.
\newline\urlprefix\url{http://dx.doi.org/10.1002/mrm.10536}

\bibitem{Xu2007}
J.~Xu, M.~Does, J.~Gore,
  \href{http://view.ncbi.nlm.nih.gov/pubmed/17374905}{Numerical study of water
  diffusion in biological tissues using an improved finite difference method},
  Physics in Medicine and Biology 52~(7).
\newline\urlprefix\url{http://view.ncbi.nlm.nih.gov/pubmed/17374905}

\bibitem{Harkins2009}
K.~D. Harkins, J.-P. Galons, T.~W. Secomb, T.~P. Trouard,
  \href{http://dx.doi.org/10.1002/mrm.22155}{Assessment of the effects of
  cellular tissue properties on {ADC} measurements by numerical simulation of
  water diffusion}, Magn. Reson. Med. 62~(6) (2009) 1414--1422.
\newline\urlprefix\url{http://dx.doi.org/10.1002/mrm.22155}

\bibitem{Russell2012}
G.~Russell, K.~D. Harkins, T.~W. Secomb, J.-P. Galons, T.~P. Trouard,
  \href{http://stacks.iop.org/0031-9155/57/i=4/a=N35}{A finite difference
  method with periodic boundary conditions for simulations of
  diffusion-weighted magnetic resonance experiments in tissue}, Physics in
  Medicine and Biology 57~(4) (2012) N35.
\newline\urlprefix\url{http://stacks.iop.org/0031-9155/57/i=4/a=N35}

\bibitem{Moroney2013}
B.~F. Moroney, T.~Stait-Gardner, B.~Ghadirian, N.~N. Yadav, W.~S. Price,
  \href{http://www.sciencedirect.com/science/article/pii/S1090780713001572}{Numerical
  analysis of {NMR} diffusion measurements in the short gradient pulse limit},
  Journal of Magnetic Resonance 234~(0) (2013) 165--175.
\newline\urlprefix\url{http://www.sciencedirect.com/science/article/pii/S1090780713001572}

\bibitem{Nguyen2014283}
D.~V. Nguyen, J.-R. Li, D.~Grebenkov, D.~L. Bihan,
  \href{http://www.sciencedirect.com/science/article/pii/S0021999114000308}{A
  finite elements method to solve the bloch–torrey equation applied to
  diffusion magnetic resonance imaging}, Journal of Computational Physics
  263~(Supplement C) (2014) 283 -- 302.
\newblock \href {http://dx.doi.org/https://doi.org/10.1016/j.jcp.2014.01.009}
  {\path{doi:https://doi.org/10.1016/j.jcp.2014.01.009}}.
\newline\urlprefix\url{http://www.sciencedirect.com/science/article/pii/S0021999114000308}

\bibitem{BELTRACHINI2015126}
L.~Beltrachini, Z.~A. Taylor, A.~F. Frangi,
  \href{http://www.sciencedirect.com/science/article/pii/S1090780715001743}{A
  parametric finite element solution of the generalised bloch–torrey equation
  for arbitrary domains}, Journal of Magnetic Resonance 259~(Supplement C)
  (2015) 126 -- 134.
\newblock \href {http://dx.doi.org/https://doi.org/10.1016/j.jmr.2015.08.008}
  {\path{doi:https://doi.org/10.1016/j.jmr.2015.08.008}}.
\newline\urlprefix\url{http://www.sciencedirect.com/science/article/pii/S1090780715001743}

\bibitem{1742-6596-490-1-012013}
D.~V. Nguyen, J.~R. Li, D.~S. Grebenkov, D.~L. Bihan,
  \href{http://stacks.iop.org/1742-6596/490/i=1/a=012013}{Modeling the
  diffusion magnetic resonance imaging signal inside neurons}, Journal of
  Physics: Conference Series 490~(1) (2014) 012013.
\newline\urlprefix\url{http://stacks.iop.org/1742-6596/490/i=1/a=012013}

\bibitem{Nguyen1080573}
V.~D. Nguyen, \href{https://www.eccomas2016.org/}{{A FEniCS-HPC framework for
  multi-compartment Bloch-Torrey models}}, Vol.~1, 2016, pp. 105--119, {QC
  20170509}.
\newline\urlprefix\url{https://www.eccomas2016.org/}

\bibitem{NGUYEN2018271}
V.-D. Nguyen, J.~Jansson, J.~Hoffman, J.-R. Li,
  \href{http://www.sciencedirect.com/science/article/pii/S0021999118305709}{A
  partition of unity finite element method for computational diffusion mri},
  Journal of Computational Physics 375 (2018) 271 -- 290.
\newblock \href {http://dx.doi.org/https://doi.org/10.1016/j.jcp.2018.08.039}
  {\path{doi:https://doi.org/10.1016/j.jcp.2018.08.039}}.
\newline\urlprefix\url{http://www.sciencedirect.com/science/article/pii/S0021999118305709}

\bibitem{NGUYEN2019176}
V.-D. Nguyen, J.~Jansson, H.~T.~A. Tran, J.~Hoffman, J.-R. Li,
  \href{http://www.sciencedirect.com/science/article/pii/S1090780719300023}{Diffusion
  mri simulation in thin-layer and thin-tube media using a discretization on
  manifolds}, Journal of Magnetic Resonance 299 (2019) 176 -- 187.
\newblock \href {http://dx.doi.org/https://doi.org/10.1016/j.jmr.2019.01.002}
  {\path{doi:https://doi.org/10.1016/j.jmr.2019.01.002}}.
\newline\urlprefix\url{http://www.sciencedirect.com/science/article/pii/S1090780719300023}

\bibitem{2019arXiv190201025L}
J.-R. {Li}, V.-D. {Nguyen}, T.~{Nguyen Tran}, J.~{Valdman}, B.~{Cong Trang},
  K.~{Van Nguyen}, V.~D. {Thach Son}, H.~A. {Tran}, H.~T.~A. {Tran}, T.~M.
  {Phuong Nguyen}, {SpinDoctor: a Matlab toolbox for diffusion MRI simulation},
  arXiv e-prints (2019) arXiv:1902.01025\href {http://arxiv.org/abs/1902.01025}
  {\path{arXiv:1902.01025}}.

\bibitem{Logg2012}
A.~Logg, K.-A. Mardal, G.~N. Wells, Automated solution of differential
  equations by the finite element method : the FEniCS book, Springer Verlag,
  2012, xIII, 723 s. : ill.

\bibitem{fenics:www}
FEniCS, Fenics project, \texttt{http://www.fenicsproject.org}.

\bibitem{FEniCSContainers}
J.~S. {Hale}, L.~{Li}, C.~N. {Richardson}, G.~N. {Wells}, Containers for
  portable, productive, and performant scientific computing, Computing in
  Science Engineering 19~(6) (2017) 40--50.
\newblock \href {http://dx.doi.org/10.1109/MCSE.2017.2421459}
  {\path{doi:10.1109/MCSE.2017.2421459}}.

\bibitem{PhysRev.104.563}
H.~C. Torrey, \href{https://link.aps.org/doi/10.1103/PhysRev.104.563}{Bloch
  equations with diffusion terms}, Phys. Rev. 104 (1956) 563--565.
\newblock \href {http://dx.doi.org/10.1103/PhysRev.104.563}
  {\path{doi:10.1103/PhysRev.104.563}}.
\newline\urlprefix\url{https://link.aps.org/doi/10.1103/PhysRev.104.563}

\bibitem{doi:10.1063/1.436751}
J.~E. Tanner, \href{https://doi.org/10.1063/1.436751}{Transient diffusion in a
  system partitioned by permeable barriers. application to nmr measurements
  with a pulsed field gradient}, The Journal of Chemical Physics 69~(4) (1978)
  1748--1754.
\newblock \href {http://arxiv.org/abs/https://doi.org/10.1063/1.436751}
  {\path{arXiv:https://doi.org/10.1063/1.436751}}, \href
  {http://dx.doi.org/10.1063/1.436751} {\path{doi:10.1063/1.436751}}.
\newline\urlprefix\url{https://doi.org/10.1063/1.436751}

\bibitem{Stejskal1965}
E.~O. Stejskal, J.~E. Tanner, \href{http://dx.doi.org/10.1063/1.1695690}{Spin
  diffusion measurements: Spin echoes in the presence of a time-dependent field
  gradient}, The Journal of Chemical Physics 42~(1) (1965) 288--292.
\newline\urlprefix\url{http://dx.doi.org/10.1063/1.1695690}

\bibitem{Does2003}
M.~D. Does, E.~C. Parsons, J.~C. Gore, Oscillating gradient measurements of
  water diffusion in normal and globally ischemic rat brain, Magn. Reson. Med.
  49~(2) (2003) 206--215.
\newblock \href {http://dx.doi.org/10.1002/mrm.10385}
  {\path{doi:10.1002/mrm.10385}}.

\bibitem{Shemesh2016}
N.~Shemesh, S.~N. Jespersen, D.~C. Alexander, Y.~Cohen, I.~Drobnjak, T.~B.
  Dyrby, J.~Finsterbusch, M.~A. Koch, T.~Kuder, F.~Laun, M.~Lawrenz,
  H.~Lundell, P.~P. Mitra, M.~Nilsson, E.~Özarslan, D.~Topgaard, C.-F. Westin,
  \href{https://onlinelibrary.wiley.com/doi/abs/10.1002/mrm.25901}{Conventions
  and nomenclature for double diffusion encoding nmr and mri}, Magnetic
  Resonance in Medicine 75~(1) (2016) 82--87.
\newblock \href
  {http://arxiv.org/abs/https://onlinelibrary.wiley.com/doi/pdf/10.1002/mrm.25901}
  {\path{arXiv:https://onlinelibrary.wiley.com/doi/pdf/10.1002/mrm.25901}},
  \href {http://dx.doi.org/10.1002/mrm.25901} {\path{doi:10.1002/mrm.25901}}.
\newline\urlprefix\url{https://onlinelibrary.wiley.com/doi/abs/10.1002/mrm.25901}

\bibitem{Dhital2019}
B.~Dhital, M.~Reisert, E.~Kellner, V.~G. Kiselev,
  \href{http://www.sciencedirect.com/science/article/pii/S1053811919300151}{Intra-axonal
  diffusivity in brain white matter}, NeuroImage 189 (2019) 543 -- 550.
\newblock \href
  {http://dx.doi.org/https://doi.org/10.1016/j.neuroimage.2019.01.015}
  {\path{doi:https://doi.org/10.1016/j.neuroimage.2019.01.015}}.
\newline\urlprefix\url{http://www.sciencedirect.com/science/article/pii/S1053811919300151}

\bibitem{Novikov2019}
D.~S. Novikov, E.~Fieremans, S.~N. Jespersen, V.~G. Kiselev,
  \href{https://onlinelibrary.wiley.com/doi/abs/10.1002/nbm.3998}{Quantifying
  brain microstructure with diffusion {MRI}: Theory and parameter estimation},
  NMR in Biomedicine 32~(4) (2019) e3998.
\newblock \href {http://dx.doi.org/10.1002/nbm.3998}
  {\path{doi:10.1002/nbm.3998}}.
\newline\urlprefix\url{https://onlinelibrary.wiley.com/doi/abs/10.1002/nbm.3998}

\bibitem{Henriques2019}
R.~N. Henriques, S.~N. Jespersen, N.~Shemesh,
  \href{https://onlinelibrary.wiley.com/doi/abs/10.1002/mrm.27606}{Microscopic
  anisotropy misestimation in spherical-mean single diffusion encoding mri},
  Magnetic Resonance in Medicine 81~(5) (2019) 3245--3261.
\newblock \href
  {http://arxiv.org/abs/https://onlinelibrary.wiley.com/doi/pdf/10.1002/mrm.27606}
  {\path{arXiv:https://onlinelibrary.wiley.com/doi/pdf/10.1002/mrm.27606}},
  \href {http://dx.doi.org/10.1002/mrm.27606} {\path{doi:10.1002/mrm.27606}}.
\newline\urlprefix\url{https://onlinelibrary.wiley.com/doi/abs/10.1002/mrm.27606}

\bibitem{TOPGAARD201798}
D.~Topgaard,
  \href{http://www.sciencedirect.com/science/article/pii/S1090780716302701}{Multidimensional
  diffusion mri}, Journal of Magnetic Resonance 275 (2017) 98 -- 113.
\newblock \href {http://dx.doi.org/https://doi.org/10.1016/j.jmr.2016.12.007}
  {\path{doi:https://doi.org/10.1016/j.jmr.2016.12.007}}.
\newline\urlprefix\url{http://www.sciencedirect.com/science/article/pii/S1090780716302701}

\bibitem{2008IJNME73361Y}
Z.~{Yuan}, J.~{Fish}, {Toward realization of computational homogenization in
  practice}, International Journal for Numerical Methods in Engineering 73
  (2008) 361--380.
\newblock \href {http://dx.doi.org/10.1002/nme.2074}
  {\path{doi:10.1002/nme.2074}}.

\bibitem{BashariRad2017}
B.~Bashari~Rad, H.~Bhatti, M.~Ahmadi, An introduction to docker and analysis of
  its performance, IJCSNS International Journal of Computer Science and Network
  Security 173 (2017) 8.

\bibitem{10.1371/journal.pone.0177459}
G.~M. Kurtzer, V.~Sochat, M.~W. Bauer,
  \href{https://doi.org/10.1371/journal.pone.0177459}{Singularity: Scientific
  containers for mobility of compute}, PLOS ONE 12~(5) (2017) 1--20.
\newblock \href {http://dx.doi.org/10.1371/journal.pone.0177459}
  {\path{doi:10.1371/journal.pone.0177459}}.
\newline\urlprefix\url{https://doi.org/10.1371/journal.pone.0177459}

\bibitem{NAP25303}
E.~National Academies~of Sciences, Medicine,
  \href{https://www.nap.edu/catalog/25303/reproducibility-and-replicability-in-science}{Reproducibility
  and Replicability in Science}, The National Academies Press, Washington, DC,
  2019.
\newblock \href {http://dx.doi.org/10.17226/25303} {\path{doi:10.17226/25303}}.
\newline\urlprefix\url{https://www.nap.edu/catalog/25303/reproducibility-and-replicability-in-science}

\bibitem{MELENK1996289}
J.~Melenk, I.~Babuška,
  \href{http://www.sciencedirect.com/science/article/pii/S0045782596010870}{The
  partition of unity finite element method: Basic theory and applications},
  Computer Methods in Applied Mechanics and Engineering 139~(1) (1996) 289 --
  314.
\newblock \href
  {http://dx.doi.org/https://doi.org/10.1016/S0045-7825(96)01087-0}
  {\path{doi:https://doi.org/10.1016/S0045-7825(96)01087-0}}.
\newline\urlprefix\url{http://www.sciencedirect.com/science/article/pii/S0045782596010870}

\bibitem{BenjaminKehlet}
J.~R. G. N.~W. Benjamin~Kehlet, Anders~Logg, Fenics project,
  \url{https://bitbucket.org/fenics-project/mshr/} (2019).

\bibitem{Logg:2010:DAF:1731022.1731030}
A.~Logg, G.~N. Wells, \href{http://doi.acm.org/10.1145/1731022.1731030}{Dolfin:
  Automated finite element computing}, ACM Trans. Math. Softw. 37~(2) (2010)
  20:1--20:28.
\newblock \href {http://dx.doi.org/10.1145/1731022.1731030}
  {\path{doi:10.1145/1731022.1731030}}.
\newline\urlprefix\url{http://doi.acm.org/10.1145/1731022.1731030}

\bibitem{geuzaine09gmsh}
C.~Geuzaine, J.~F. Remacle, {Gmsh: a three-dimensional finite element mesh
  generator with built-in pre- and post-processing facilities}, International
  Journal for Numerical Methods in Engineering.

\bibitem{salome}
A.~{Ribes}, C.~{Caremoli}, Salomé platform component model for numerical
  simulation, in: 31st Annual International Computer Software and Applications
  Conference (COMPSAC 2007), Vol.~2, 2007, pp. 553--564.
\newblock \href {http://dx.doi.org/10.1109/COMPSAC.2007.185}
  {\path{doi:10.1109/COMPSAC.2007.185}}.

\bibitem{ansa}
Beta cae systems, ansa pre-processor: The advanced cae pre-processing software
  for complete model build up., \url{https://www.beta-cae.com}.

\bibitem{meshio}
N.~Schl\"omer, meshio, \url{https://github.com/nschloe/meshio} (2019).

\bibitem{colab}
G.~Inc., Google colaboratory, \url{https://github.com/jupyter/colaboratory}
  (2014).

\bibitem{matplotlib}
J.~H. et~al., Matplotlib.

\bibitem{paraview}
K.~I. Sandia~Corporation, Paraview.

\bibitem{GREBENKOV2010181}
D.~S. Grebenkov,
  \href{http://www.sciencedirect.com/science/article/pii/S1090780710001199}{Pulsed-gradient
  spin-echo monitoring of restricted diffusion in multilayered structures},
  Journal of Magnetic Resonance 205~(2) (2010) 181 -- 195.
\newblock \href {http://dx.doi.org/https://doi.org/10.1016/j.jmr.2010.04.017}
  {\path{doi:https://doi.org/10.1016/j.jmr.2010.04.017}}.
\newline\urlprefix\url{http://www.sciencedirect.com/science/article/pii/S1090780710001199}

\bibitem{doi:10.1002/cne.23181}
J.~P. Kinney, J.~Spacek, T.~M. Bartol, C.~L. Bajaj, K.~M. Harris, T.~J.
  Sejnowski,
  \href{https://onlinelibrary.wiley.com/doi/abs/10.1002/cne.23181}{Extracellular
  sheets and tunnels modulate glutamate diffusion in hippocampal neuropil},
  Journal of Comparative Neurology 521~(2) (2013) 448--464.
\newblock \href
  {http://arxiv.org/abs/https://onlinelibrary.wiley.com/doi/pdf/10.1002/cne.23181}
  {\path{arXiv:https://onlinelibrary.wiley.com/doi/pdf/10.1002/cne.23181}},
  \href {http://dx.doi.org/10.1002/cne.23181} {\path{doi:10.1002/cne.23181}}.
\newline\urlprefix\url{https://onlinelibrary.wiley.com/doi/abs/10.1002/cne.23181}

\bibitem{Carim-Todd678}
L.~Carim-Todd, K.~G. Bath, G.~Fulgenzi, S.~Yanpallewar, D.~Jing, C.~A. Barrick,
  J.~Becker, H.~Buckley, S.~G. Dorsey, F.~S. Lee, L.~Tessarollo,
  \href{http://www.jneurosci.org/content/29/3/678}{Endogenous truncated trkb.t1
  receptor regulates neuronal complexity and trkb kinase receptor function in
  vivo}, Journal of Neuroscience 29~(3) (2009) 678--685.
\newblock \href
  {http://arxiv.org/abs/http://www.jneurosci.org/content/29/3/678.full.pdf}
  {\path{arXiv:http://www.jneurosci.org/content/29/3/678.full.pdf}}, \href
  {http://dx.doi.org/10.1523/JNEUROSCI.5060-08.2009}
  {\path{doi:10.1523/JNEUROSCI.5060-08.2009}}.
\newline\urlprefix\url{http://www.jneurosci.org/content/29/3/678}

\bibitem{scipy}
S.~Developers, Scipy, \url{https://www.scipy.org} (2001).

\bibitem{Nitsche1971}
J.~Nitsche, \href{https://doi.org/10.1007/BF02995904}{{\"U}ber ein
  variationsprinzip zur l{\"o}sung von dirichlet-problemen bei verwendung von
  teilr{\"a}umen, die keinen randbedingungen unterworfen sind}, Abhandlungen
  aus dem Mathematischen Seminar der Universit{\"a}t Hamburg 36~(1) (1971)
  9--15.
\newblock \href {http://dx.doi.org/10.1007/BF02995904}
  {\path{doi:10.1007/BF02995904}}.
\newline\urlprefix\url{https://doi.org/10.1007/BF02995904}

\bibitem{GAMM:GAMM201490018}
P.~Hansbo, \href{http://dx.doi.org/10.1002/gamm.201490018}{Nitsche's method for
  interface problems in computational mechanics}, GAMM-Mitteilungen 28~(2)
  (2005) 183--206.
\newblock \href {http://dx.doi.org/10.1002/gamm.201490018}
  {\path{doi:10.1002/gamm.201490018}}.
\newline\urlprefix\url{http://dx.doi.org/10.1002/gamm.201490018}

\end{thebibliography}

\end{document}